\documentclass[10pt]{article}
\usepackage{amssymb,latexsym}
\usepackage{amsmath,amsbsy}
\usepackage{epsfig}
\usepackage{feynmf}
\DeclareMathOperator{\pperp}{\mathbf{p}^{\perp}}
\DeclareMathOperator{\Pperp}{\mathbf{P}^{\perp}}

\DeclareMathOperator{\Pplus}{P^{+}}
\DeclareMathOperator{\pplus}{p^{+}}
\DeclareMathOperator{\qperp}{\mathbf{q}^{\perp}}

\DeclareMathOperator{\kperp}{\mathbf{k}^{\perp}}
\DeclareMathOperator{\kplus}{k^{+}}
\DeclareMathOperator{\kaperp}{\mathbf{\kappa}^{\perp}}

\DeclareMathOperator{\Pee}{P^{\mu}}
\DeclareMathOperator{\Pp}{{P}^{\prime}{}^{\mu}}
\DeclareMathOperator{\Pbar}{\bar{P}^{\mu}}
\DeclareMathOperator{\Pbarplus}{\bar{P}^{+}}

\DeclareMathOperator{\ndpd}{\mathcal{F}(X, \zeta, t)}
\DeclareMathOperator{\ie}{i \epsilon}
\DeclareMathOperator{\Res}{Res}
\DeclareMathOperator{\ofpd}{F(x,\xi,t)}
\DeclareMathOperator{\dw}{D_{\text{W}}}
\DeclareMathOperator{\bigdw}{\Delta_{\text{W}}}
\DeclareMathOperator{\mn}{\langle \mu^2 \rangle}
\DeclareMathOperator{\pp}{p^{\prime}}
\DeclareMathOperator{\GDA}{\Phi(z, \zeta, s)}
\DeclareMathOperator{\lam}{\Lambda}

\begin{document}

\title{\hskip7cm NT@UW 01-22\\
  Exploring Skewed Parton Distributions with Two-body Models on the Light Front II: covariant Bethe-Salpeter approach}
\author{ B.~C.~Tiburzi\footnote{Email: bctiburz@u.washington.edu} \; and G.~A.~Miller\\
        Department of Physics\\ 
University of Washington\\      
Box $351560$\\  
Seattle, WA $98195-1560$ }
\date{\today}
\maketitle

\begin{abstract}
We explore skewed parton distributions for two-body, light-front wave functions. In order to access all 
kinematical r\'{e}gimes, we adopt a covariant Bethe-Salpeter approach, which makes use of the underlying 
equation of motion (here the Weinberg equation) and its Green's function. Such an approach allows for the 
consistent treatment of the non-wave function vertex (but rules out the case of phenomenological wave functions 
derived from \emph{ad hoc} potentials). Our investigation centers around checking internal consistency 
by demonstrating time-reversal invariance and continuity between valence and non-valence r\'egimes.
We derive our expressions by assuming the effective $qq$ potential is independent of the mass squared,
and verify the sum rule in a non-relativistic approximation in which the potential is 
energy independent. We consider bare-coupling as well as interacting 
skewed parton distributions and develop approximations for the Green's function which preserve 
the general properties of these distributions. Lastly we apply our approach to time-like
form factors and find similar expressions for the related generalized distribution amplitudes. 
\end{abstract}

\section{Introduction}

Recently there has been considerable interest in the connection between hard inclusive and exclusive reactions, which has
been, in part, due to the unifying r\^{o}le of skewed parton distributions (SPD's) \cite{Muller:1994fv, Ji:1997ek, Radyushkin:1997ki}. 
Aside from being the natural marriage of form factors and parton distributions, SPD's appear when one tries to calculate scattering 
amplitudes such as those for Compton scattering \cite{Radyushkin:1997ki,Ji:1998xh, Collins:1999be}, the electroproduction of 
mesons \cite{Radyushkin:1996ru, Collins:1997fb}, and di-jet production by pions on nuclei \cite{Frankfurt:2000jm}. 

There is much effort underway related to the measurement of these functions \cite{Saull:1999kt}. Intuitively clear, but simple
models \cite{Radyushkin:1997ki,Ji:1997gm,Diehl:1999kh} have been used to provide first calculations 
(more sophisticated approaches have been pursued \cite{Burkardt:2000uu,Brodsky:2001xy, Petrov:1998kf}) and physical interpretations
have been elucidated in the forward limit \cite{Burkardt:2000za}. In this paper, we attempt to gain intuition about the 
structure of these distributions by considering two-body, light-front models 
as an explicitly worked example. Continuing with our previous work \cite{Tiburzi:2001ta}, we consider the SPD for a toy scalar \emph{meson} 
consisting of two, equally massive scalar \emph{quarks}. The features which are 
general in our approach are the energy denominators, which will be encountered in any calculation of SPD's (e.g.  for a hadron of 
spin-$\frac{1}{2}$ consisting of spin-$\frac{1}{2}$ quarks). Simple two-body, wave function models of SPD's suffer from an inability 
to describe all kinematical r\'egimes physically accessible. On the other hand, using a light-front Fock-space decomposition of the 
initial and final hadron states gives one a powerful general expression for the SPD's as sums of diagonal and non-diagonal 
overlaps \cite{Diehl:2001xz} encompassing all kinematical r\'egimes. In practice, however,
this approach is limited by our current knowledge of the higher Fock-space components of hadrons and one seems constrained to use 
Gaussian \emph{Ans\"{a}tze} for these components \cite{Diehl:1999kh} for which essential 
physics may be absent. Generalizing a rather prescient method due to Einhorn
\cite{Einhorn:1976uz}, we avoid the need to obtain higher Fock-space components explicitly
by using the Bethe-Salpeter equation, and crossing symmetry. 
At this stage, we sacrifice exactness (by neglecting the mass squared dependence of the interaction throughout) in order to
build intuition about these distributions in this approach.
We find that one need only use the lowest
Fock-space components (here the $q\bar{q}$ pair) and the equation of motion
to access all kinematical r\'egimes of the SPD with model wave functions. 
We note that recently a similar analysis, although tentative, appeared for spin-$\frac{1}{2}$ quarks
in the context of SPD's in \cite{Choi:2001fc}.

We begin in section \ref{CBS} by writing the covariant Bethe-Salpeter
equation for our meson and extract from this equation the light front wave function 
and its equation of motion (the Weinberg equation). By way of orthogonality and completeness, the
Green's function for the Weinberg equation is then defined. In section \ref{DVCS}
we apply this analysis to the covariant, leading-twist diagram for 
deeply virtual Compton scattering (DVCS). We carefully demonstrate that 
in the deeply virtual limit, SPD's can be calculated from the light-front time-ordered 
triangle diagrams for the form factor, making clear the close connection between the sum
rule and light-front time-ordered perturbation theory.  These diagrams are evaluated in section
\ref{TD} in terms of $qq$ light front wave functions where the Z-graph is
dealt with by crossing the interaction. The resulting bare-coupling SPD's are derived 
and their simplicity is used to discuss continuity, $x$-symmetry and time reversal invariance properties.
The demonstration of time reversal invariance for the bare-coupling SPD's is started in section \ref{TD}
and completed in Appendix A. 

The bare-coupling SPD's are only (partially) valid at high $|t|$, 
which is unfortunately not likely accessible in experiments. Moreover, the effects of higher Fock components
in this approach are contained in the photon vertex function. Thus inclusion of interactions at the photon
vertex is required and carried out in section \ref{T}, where we derive the photon vertex function 
in terms of the interacting theory Green's function. Finally the general form of the SPD's is deduced 
in section \ref{AGAIN}. These distributions satisfy the necessary symmetry properties and are continuous 
due to the behavior of the Green's function near the end points. As calculation of the Green's function
is likely burdensome, we discuss in section \ref{APPROX} possible approximations to the Green's function which 
maintain the symmetry and continuity of the distributions. We explore first the Born approximation 
valid at high $|t|$ (which does not result in the bare-coupling SPD's). Next we examine the non-relativistic 
scheme and finally the closure approximation. 

The stage is set to apply the non-relativistic approximation scheme to the Wick-Cutkosky model 
in the weak-binding limit (section \ref{WFNS}). We use this example to illustrate how to cross the interaction
for partons moving backwards through time and thereby explicate that in general an \emph{ad hoc} potential (which might 
generate some phenomenological hadronic wave function) lacks the field theoretic nature to accommodate such crossing.  
We then verify the sum rule for the form factor in the non-relativistic scheme, which is accomplished by using only 
the valence contribution to the SPD. Appendix B makes explicit the recovery of the Drell-Yan-West formula
for the form factor in the limit of zero skewness. Furthermore, we comment there on the general consequences and limitations 
of approximating the interaction to be independent of the mass squared. 

The extension to time-like form factors is carried out in Appendix C. We proceed analogously with our consideration of 
SPD's by deriving the sum rule for time like form factors in terms of generalized distribution amplitudes. Then we calculate
the generalized distribution amplitudes in terms of the relevant light-front time-ordered graphs. Lastly we conclude briefly in 
section \ref{conc}. 

\section{Light-front reduction of the Bethe-Salpeter equation} \label{CBS}
\begin{figure}
\begin{center}
\begin{fmffile}{fmfBS}
	\parbox{20mm}{
	\begin{fmfchar*}(20,30)
	\fmfleft{a,b,c}
	\fmf{double,label=$R$,label.side=left}{b,in}
	\fmf{plain,left=.2,label=$p$,label.side=left}{in,f}
 	\fmf{plain,right=.2}{in,d}
	\fmfright{dd,d,f,ff} 
    	\fmfv{decor.shape=circle,decor.filled=full,decor.size=.3w}{in}
	\end{fmfchar*}} =
	\parbox{40mm}{  	
	\begin{fmfchar*}(40,30)
	\fmfleft{a,b,c}
	\fmf{double,tension=1,label=$R$,label.side=left}{b,in}
	\fmfpolyn{empty,tension=.5,label=$V$}{Z}{4}
	\fmf{plain,right=.2,tension=.5}{in,Z4}
        \fmf{plain,left=.2,tension=.5,label=$k$,label.side=left}{in,Z3}
	\fmf{plain,right=.1}{Z1,d}
    	\fmf{plain,left=.1,label=$p$,label.side=left}{Z2,f}
 	\fmfright{dd,d,f,ff}
    	\fmfv{decor.shape=circle,decor.filled=full,decor.size=.15w}{in}
    	\end{fmfchar*}}
\end{fmffile}
\end{center}
\caption{Diagrammatic representation of the Bethe-Salpeter equation. The blob represents the vertex function $\Gamma$.}
\label{fBS}
\end{figure}
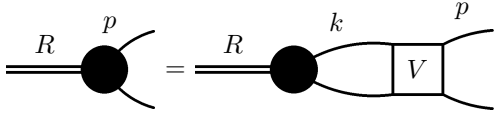

We start by writing down the equation for the meson vertex function $\Gamma$. It satisfies a simple Bethe-Salpeter
equation, first schematically as operators (see Figure \ref{fBS}):
\begin{equation} 
\Gamma = - V G_{o} \Gamma,
\end{equation}
where $G_{o}$ is the full, renormalized, two-particle Green's function (which can be expressed as a product
of two, renormalized, single-particle propagators). Here $V$ is the interaction potential, which is the negative
of the irreducible two-to-two scattering kernel. Let us denote the meson four-momentum by $R$,
the quark's by $p$, and the physical masses by $R^2 \equiv M^2, p^2 \equiv m^2$.  
Since our quarks are scalar particles, the renormalized, single particle propagator has a Klein-Gordon form:
\begin{equation}
G_{o}(k) = \frac{i}{(k^2 - m^2)[1+ (k^2 - m^2)f(k^2 -m^2)]+ i \epsilon},
\end{equation}
where the residue is one near the physical mass pole and the function $f(k^2-m^2)$ characterizes the renormalized, one-particle 
irreducible self interactions. It is a matter of later notational convenience that we write $f = f(k^2 - m^2)$. 
We shall assume there are no poles (besides at the physical mass) in the propagator. This is consistent with model
studies \cite{Roberts:1994dr}. In the momentum representation, the Bethe-Salpeter equation then appears as 
\begin{equation} \label{BS}
\Gamma(p,R) = - i \int \frac{d^4k}{(2\pi)^4} \Psi_{BS}(k,R) V(k,p),
\end{equation}
in which we have defined the Bethe-Salpeter wave function $\Psi_{BS}$ as
\begin{equation}\label{psiBS}
\Psi_{BS}(k,R) = G_{o}(k) \Gamma(k,R) G_{o}(R-k).
\end{equation}

The Bethe-Salpeter wave function is covariant, thus in order to perform the light front 
reduction we must project out the $x^{+} = 0$ hypersurface where the initial conditions,
wave functions, \emph{etc.} of our system are defined \cite{Brodsky:1985vp}. (We define the plus and minus components 
of any vector by $x^{\pm} = (x^{0} \pm x^{3})/\sqrt{2}$.) This projection onto the initial 
surface gives us the light front wave function, \emph{vis} $\Psi_{LF} = \int dk^{-} \Psi_{BS}$.
The remarkable property of light front wave functions is that they depend not on the total momentum 
of the system but on the relative momenta of the constituents \cite{Dirac:1949cp}. Since the constituents 
of our meson are two equally massive quarks, the relative momenta take a simple form. Writing this dependence
out explicitly, we see
\begin{equation}\label{lfwf}
\Psi_{LF} (k, R) \equiv \frac{\pi i}{R^{+} x (1-x)} \psi \big(x, \kperp_{\text{rel}} = \kperp - x \mathbf{R}^{\perp} \big), 
\end{equation}
with $\kplus /R^{+} \equiv x$ and the pre-factors chosen with malice aforethought. In order to perform the $x^{+} = 0$ projection
on the Bethe-Salpeter wave function, we must know something about the analytic structure of the vertex function. We would like to 
perform the $k^-$ integration simply by assuming there are neither $k^-$ poles, nor branch cuts in the vertex function. 
Looking at the effective $qq$ potential in Eq. \eqref{BS}, shows that the $p^-$ dependence of the interaction $V(k,p)$ gives rise to 
the $p^{-}$ poles of the vertex function $\Gamma(p,R)$.  Thus taking the effective $qq$ potential $V(k,p)$ to be independent 
of the minus momenta eliminates the vertex function's dependence on the light-front energy and allows us to easily perform the light-front
projection. The effective two-body interaction in $1+1$ QCD, for example, is independent of the minus momentum. Though certainly this 
independence is far from general. Calculating $V(k,p)$ from light-front, time-ordered graphs, however, requires the \emph{on shell} 
condition which removes any functional dependence on the minus momenta from $V$ and hence $\Gamma$.
Now carrying out the $k^-$ integration on the Bethe-Salpeter wave function, we find
\begin{align} \label{wf}
\psi(x, \kperp_{ \text{rel} } ) & = \frac{R^{+} x (1-x)}{\pi i} \int dk^{-} \Psi_{BS} (k, R) \notag \\
			     & = \bigdw (x,\kperp_{\text{rel}}|M^2) \Gamma (x,\kperp_{ \text{rel} } ),
\end{align}
where we have defined
\begin{equation}
\bigdw(x,\kperp_{\text{rel}}|M^2) = \frac{\dw(x,\kperp_{\text{rel}}|M^2)}
{1+ x \dw^{-1}(x,\kperp_{\text{rel}}|M^2) f\Big( x \dw^{-1}(x,\kperp_{\text{rel}}|M^2)\Big) }
\end{equation}
with the help of the abbreviation $\dw$ for the Weinberg propagator
\begin{equation}
\dw(x,\kperp_{\text{rel}}|M^2) = \frac{1}{M^2 - \frac{m^2 - \kperp_{\text{rel}}^2}{x(1-x)}}.
\end{equation}
The projection carried out in equation \ref{wf} can also be utilized to establish the symmetry relation
\begin{equation} \label{symm}
x \; f\Big( x \dw(x,\kperp|M^2)\Big) = (1-x) f\Big( (1-x)\dw(1-x,\kperp|M^2) \Big),
\end{equation}
required for a system with equally massive quarks. 

Notice we have implicitly altered the functional dependence from $\Gamma = \Gamma (k, R)$ 
to $\Gamma = \Gamma (x, \kperp_{\text{rel}})$. As a result of the 
integration in Eq. \eqref{wf}, $x$ is restricted: $x \in (0,1)$. Thus the light-front wave function vanishes for $x$ outside this range. 
Eq. (\ref{wf}) gives the  relation between the vertex function and the light front wave function. 
This relation will be exploited to evaluate Feynman graphs in terms of wave functions below---provided, of course, $x \in (0,1)$.

It remains to find the equation of motion satisfied by the wave function $\psi(x, \kperp_{ \text{rel} } )$.
We accomplish this by performing the $k^{-}$ integration in the Bethe-Salpeter equation (\ref{BS}).
\begin{align} \label{wein}
\bigdw^{-1}(x,\kperp_{\text{rel}}|M^2) \psi(x, \kperp_{ \text{rel} } ) & =  - i  \int \frac{d^4p}{(2\pi)^4} 
	\Psi_{BS}(p,R) V(p,k) \notag \\
	& =  \int \frac{dy d\pperp_{ \text{rel} }} {2 (2 \pi)^3 y (1-y)} V(y, \pperp_{\text{rel}}; x, \kperp_{\text{rel}}) 
	\psi(y, \pperp_{\text{rel}}),
\end{align}
with $y \equiv \pplus/R^{+}$, and $\pperp_{\text{rel}} = \pperp - y \mathbf{R}^{\perp}$. The above equation can be written as the
Weinberg equation \cite{Weinberg:1966jm} by relocating the renormalized self interactions into the definition of the potential. 
To do so, we write the above equation more suggestively as an eigenvalue equation: 
$(M_{n}^{2} - \hat{H})\psi_{n} = 0$, where $\hat{H} = \hat{H_{o}} + \hat{V}$ with $\hat{H_{o}}$ as the kinetic term 
$\frac{ \kperp_{ \text{rel} }^2 + m^2 } {x (1-x)}$ and $\hat{V}$ as the potential, whose action on $\psi$ is spelled out by
\begin{equation}
\hat{V} \psi(x,\kperp) = \int \frac{dy d\pperp}{2 (2\pi)^3 y(1-y)} \Bigg[ V(x,\kperp;y,\pperp) - 
2(2\pi)^3 \delta(x-y)\delta^2(\kperp -\pperp)\lam(x,\kperp) \Bigg] \psi(y,\pperp),
\end{equation}
with
\begin{equation}
\lam(x,\kperp) = x \dw^{-2}(x,\kperp|M^2) f\Big( x \dw^{-1}(x,\kperp|M^2)\Big).
\end{equation}
At this level, the action of the potential is still exact, however, the self interactions as well as the potential function
depend on $M^{2}_{n}$. If we neglect this dependence (as is possible in $1+1$ QCD, or to leading order in a weakly bound system
where $M^{2}_{n} \approx 4 m^2$) then the Weinberg equation becomes a standard eigenvalue equation. Under this assumption, the
eigenfunctions $ \{ \psi_{n} \}$ 
satisfy orthonormality and completeness, which appear in our normalization convention as
\begin{equation} \label{ortho}
\int \frac{dx d\kperp}{2 (2 \pi)^3 x(1-x)} \psi_{n}(x, \kperp) \psi^{*}_{m}(x, \kperp) = \delta_{n m},
\end{equation}
\begin{equation} \label{compl}
\sum_{n} \psi_{n}(x, \kperp) \psi^{*}_{n}(y, \pperp) = 2 (2 \pi)^3 x(1-x) \delta(x - y) \delta^2(\kperp - \pperp).
\end{equation} 

The Green's function for the Weinberg equation 
\begin{equation} \label{green}
G(x, \kperp; y,\pperp | R^2) = \sum_{n} \frac{\psi_{n}(x, \kperp) \psi_{n}^{*}(y, \pperp)}{R^2 - M_{n}^2 + \ie}
\end{equation}
thus satisfies the equation 
\begin{equation} \label{greeneq}
\Big( R^2 - \hat{H} \Big) G(x, \kperp; y,\pperp | R^2) = 2 (2 \pi)^3 x(1-x) \delta(x-y) \delta^2(\kperp - \pperp).
\end{equation}
Clearly approximating away the $M^2$ dependence of the interaction has limitations. We spell out some of the relevant 
consequences in Appendix B. 

\section{Deeply virtual Compton scattering} \label{DVCS}

\begin{figure}
\begin{center}
\begin{fmffile}{fmfdvcs}
\begin{fmfchar*}(40,40)
  \fmfleft{em,ep} \fmflabel{$q$}{ep} \fmflabel{$P$}{em}
  \fmf{heavy}{em,Zee}
  \fmf{photon}{ep,Zip}
  \fmf{fermion,label=$k$,label.side=left}{Zee,Zip}
  \fmf{fermion,label=$k+q$,label.side=left}{Zip,Zam}				 
  \fmf{quark,tension=.1,label=$P-k$,label.side=right}{Zee,Zff}
  \fmf{quark,label=$k+\Delta$,label.side=left}{Zam,Zff}
  \fmf{photon}{Zam,f}	
  \fmf{heavy}{Zff,fb}
  \fmfright{fb,f} \fmflabel{$P^{\prime}$}{fb} \fmflabel{$q^{\prime}$}{f}
  \fmfdot{Zip,Zam}
  \fmfv{decor.shape=circle,decor.filled=full,decor.size=.1w}{Zee,Zff}
\end{fmfchar*}
\end{fmffile}
\caption{The leading DVCS diagram}
\label{dvcs}
\end{center}
\end{figure}
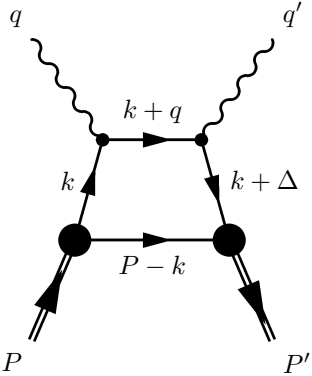

Skewed parton distributions show up, for example, as generalized form factors in deeply virtual Compton scattering. We will introduce 
the SPD's in this way. In our toy model, we imagine a virtual photon interacting with our meson (labeled as $\pi$) but leaving it 
intact $P^2 = P^{\prime}{}^2 = M^2$:
\begin{equation}
\gamma^{*}(q) + \pi(P) \longrightarrow \gamma(q^{\prime}) + \pi(P^{\prime}) \notag.
\end{equation}
This process is certainly not imaginable in any realistic experimental setup, however, it is related by crossing symmetry to the 
physical reaction $\gamma^{*} + \gamma \to \pi + \pi$ which was investigated in \cite{Diehl:1998dk}. The relation between SPD's 
and the generalized distribution amplitudes (GDA's)---which characterize two pion production---was pursued in \cite{Polyakov:1999td}. 
In Appendix C, we illustrate this connection in our light-front approach by calculating our model's GDA via the time-like form factor.

Returning to Compton scattering, the momentum transfer suffered by the meson we denote by $\Delta^{\mu} \equiv \Pp - \Pee$ 
(and its invariant $t \equiv \Delta^2$), where the $\mu$ labels $(-,+,1,2)$.
The leading-twist (\emph{handbag}) diagram for this process is depicted in Figure \ref{dvcs} where we give the parameterization of
the intermediate momenta. We follow the standard simplification \cite{Brodsky:2001xy} and choose a frame where $q^{+} = 0$ \cite{pairs}, 
so that
$q^2 = -\mathbf{q}^{\perp}{}^2 \equiv - Q^2$. In the deeply virtual limit, $Q^2$ is taken to be large while the ratio 
$\frac{Q^2}{2 P \cdot q} \equiv \zeta$ is finite. This defines $\zeta$ as the Bjorken variable for deeply virtual Compton scattering. 
In this section, we work in the asymmetrical frame where the initial meson's transverse momentum vanishes and consequently
$q^{-} = Q^2 / 2 \zeta \Pplus$. Lastly, we define $X$ as the ratio of the active quark's plus-momentum to that of the initial 
meson: $X = \kplus/\Pplus$. 

At leading twist, we can ignore interactions at the quark-photon vertex as well as quark self-interactions between photon absorbtion
and emission. Evaluating the diagram shown in the figure thus yields
\begin{equation}  \label{diag}
M^{\mu\nu}_{A} = i \int \frac{d^4k}{(2 \pi)^4} \frac{(2 k^{\mu} + q^{\mu})(2 k^{\nu} +  \Delta^{\nu} + q^{\nu})} 
{(k + q)^2 - m^2 + \ie} G_{o}(k)\Gamma(k,P)G_{o}(k+\Delta)\Gamma^{*}(k+ \Delta,P + \Delta)G_{o}(P-k) , 
\end{equation}
where $A$ signifies that this contribution is that of Figure \ref{dvcs}. We also need to calculate the crossed 
diagram $M^{\mu\nu}_{B}$ to get the covariant amplitude $M^{\mu\nu}$.

The denominator in (\ref{diag}) which depends on the momentum $q$ can be dramatically simplified
in the deeply virtual limit. Furthermore there is a cancellation in this limit from the numerator of $M^{-+}_{A}$:
\begin{equation}
\frac{2k^{-} + q^{-}}{(k + q)^2 - m^2 + \ie } \to \frac{1}{2 \Pplus}\frac{1}{X - \zeta + \ie}
\end{equation}
There is an analogous cancellation in the deeply virtual limit when we evaluate the crossed diagram $M^{-+}_{B}$.
\begin{equation}
\frac{2 k^{-} + \Delta^{-} + q^{-}}{(k + \Delta - q)^2 -m^2 + \ie} \to \frac{1}{2 \Pplus} \frac{1}{X - \ie}
\end{equation}  
The remaining terms in the integral are the same for each diagram. Thus in the deeply virtual limit, we can decompose the 
Compton amplitude as
\begin{equation} \label{decomp}
M^{-+} \approx  - \int_{0}^{1}  (2 X - \zeta) \ndpd 
\Big( \frac{1}{X - i \epsilon} + \frac{1}{X - \zeta + i \epsilon} \Big) dX.
\end{equation}

Removing the $d\kplus$ integral from the expressions for $M_{A}^{-+}$ and $M_{B}^{-+}$ reveals 
\begin{equation} \label{covartri}
\ndpd = - i \int \frac{dk^{-}d\kperp}{(2\pi)^4} G_{o}(k) \Gamma(k,P) G_{o}(k+\Delta)\Gamma^{*}(k+\Delta,P+\Delta) G_{o}(P-k) .
\end{equation}
This is just the integrand of the covariant triangle diagram which gives the electromagnetic
form factor. In this way we rediscover the sum rule of Ji \cite{Ji:1997ek}:
\begin{equation} \label{sum}
\int_{0}^{1}  \frac{2 X - \zeta}{2 - \zeta} \ndpd dX  = F(t),
\end{equation} 
where the factor $(2-\zeta)^{-1}$ appears because we normalize to the average meson momentum $\Pbar = (\Pp + \Pee)/2$.
The kinematics gives the relation $\Pbarplus = \Pplus (1 - \zeta /2)$. The other factor, $2 X - \zeta$, removes the derivative 
 coupling at the photon vertex from the SPD. The $\zeta$-independence of the form factor is due 
 to Lorentz covariance. Once we have expressed $\ndpd$ in terms of light-front wave functions, we will be concerned with 
 verifying Eq. (\ref{sum}). 

 \section{Triangle diagram and the Z-graph} \label{TD}

 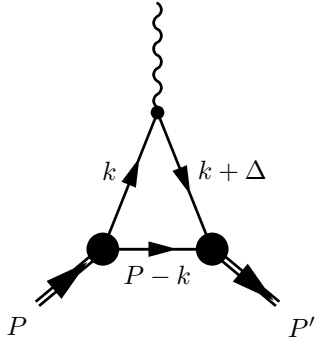
\begin{figure}
 \begin{center}
  \begin{fmffile}{fmftri}
  \begin{fmfchar*}(40,40)
     \fmfleft{em,space}  \fmflabel{$P$}{em}
    \fmf{heavy}{em,Zee}
      \fmf{fermion,tension=.4,label=$k$,label.side=left}{Zee,Zip}
      \fmf{photon}{Zip,Woo} 
      \fmftop{Woo}   
      \fmf{fermion,tension=.4,label=$k+\Delta$,label.side=left}{Zip,Zam}
      \fmf{fermion,tension=.4, label=$P-k$,label.side=right}{Zee,Zam}
    \fmf{heavy}{Zam,fb}
    \fmfright{fb,spaced} \fmflabel{$P^{\prime}$}{fb} 
    \fmfdot{Zip}
    \fmfv{decor.shape=circle,decor.filled=full,decor.size=.1w}{Zee,Zam}
  \end{fmfchar*}
  \end{fmffile}
  \caption{The triangle diagram for the electromagnetic form factor}
  \label{tri}
  \end{center}
  \end{figure}

  We have reduced determining the SPD to a calculation of the covariant triangle diagram for our meson's electromagnetic form
  factor. In the previous section, we employed a special reference frame in which the virtual photon is space-like ($q^{+} = 0$). 
  Thus in calculating the triangle graph from Eq. (\ref{covartri}), we no longer have the freedom remaining to choose $\Delta^{+} = 0$ 
  which is often done in calculating form factors. We must proceed on a course parallel to Sawicki \cite{Sawicki:1991sr}, 
  who considered such general expressions for form factors earlier.  

  The $k^{-}$ integral appearing in Eq. (\ref{covartri}) is convergent which enables evaluation by choosing to close 
  the contour in either the upper- or lower-half complex plane. Without any loss of generality, we choose $\Delta^{+} < 0$.
  The poles of our integrand are located at
  \begin{equation} \label{poles}
  \begin{cases} 
    k^{-}_{a}  & = - \Delta^{-} + \frac{m^2 + (\kperp + \Delta^{\perp})^2}{2(k^{+} + \Delta^{+})} - \frac{\ie}{2(k^{+} + \Delta^{+})}\\
    k^{-}_{b}  & = \frac{m^2 + \kperp{}^{2}}{2\kplus} - \frac{\ie}{2\kplus}\\
    k^{-}_{c}  & = P^{-} - \frac{m^2 + (\Pperp - \kperp)^2}{2(\Pplus - \kplus)} + \frac{\ie}{2(\Pplus - \kplus)}. 			
  \end{cases} 
  \end{equation}
 When $\kplus < 0$, all poles lie in the upper-half plane. On the other hand,  for $\kplus > \Pplus$ all poles
 lie in the lower-half plane. Then for both of these cases, we merely choose the contour which avoids enclosing any poles and, by Cauchy,
 the integral vanishes. To evaluate the $k^{-}$ integral for $0  < \kplus < - \Delta^{+}$ and $-\Delta^{+} < \kplus < \Pplus$, we must 
 enclose at least one pole when closing the contour. 

 First let us look at the case $- \Delta^{+} < k^{+} < \Pplus$. This is the familiar region which survives 
 when $\Delta^{+} = 0$ (of course now $\Delta^{+} = - \zeta P^{+}$). Closing the contour in the upper-half plane
 we enclose only the pole at $k_{c}^{-}$ and the integral is $+ 2 \pi i \Res(k_{c}^{-})$. Dropping pre-factors for the 
 moment, we find (see Figure \ref{tri}):
 \begin{align}
 \frac{ \Gamma(X,\kperp - X \Pperp) }{(k_{c}^{-} - k_{b}^{-})\big[ 1+ 2 \kplus (k_{c}^{-} - k_{b}^{-})
f\big(2 \kplus(k_{c}^{-} - k_{b}^{-} )\big) \big]}
 & \frac{\Gamma^{*}(Y,\kperp -Y \Pperp + (1-Y) \Delta^{\perp})) }
 {(k_{c}^{-} - k_{a}^{-})\big[ 1 + 2 (\kplus + \Delta^{+})(k_{c}^{-} - k_{a}^{-}) 
f\big( 2(\kplus + \Delta^+)(k_{c}^{-} - k_{a}^{-})   \big)   \big]}  \notag \\
 & = 4 (1-\zeta) \Pplus{}^{2} \psi(X, \kperp - X \Pperp) \psi^{*}(Y,\kperp -Y \Pperp + (1-Y) \Delta^{\perp}),  \notag
 \end{align} 
where $Y\equiv \frac{X-\zeta}{1-\zeta} > 0$. Now we merely shift the $\kperp$ integral and work in the symmetrical 
frame \cite{boost} in which $\mathbf{\bar{P}}^{\perp} = 0$. Thus for $X > \zeta$ we have
\begin{equation} \label{ndpd1}
 \ndpd = \frac{1- \zeta}{X(1-X)(X-\zeta)} \int \frac{d\kperp}{2 (2\pi)^3}  
\psi(X, \kperp) \psi^{*}(Y, \kperp + (1 - \bar{x})\Delta^{\perp}),
\end{equation}
where the average of the plus momentum arguments between the wave functions is  $\bar{x} = (X +Y)/2$. We have been employing
ratios of plus momenta with respect to the initial meson. To instructively counter this malady, we change variables from
Radyushkin's \cite{Radyushkin:1997ki} to those of Ji \cite{Ji:1997ek}.  To this end we define a new ratio of plus momenta with 
respect to the 
average meson momentum $x \equiv \kplus / \Pbarplus$, inverted this reads  $X = \frac{x + \xi}{1+ \xi}$. Furthermore
$\zeta = \frac{2 \xi}{1+\xi}$. The new symmetrical SPD (for $x>\xi$) is just 
$\ofpd = (1 + \xi) \mathcal{F}\big(X(x,\xi),\zeta(\xi),t \big)$ and 
absorbs the Jacobian factor lurking in the sum rule \eqref{sum}, \emph{vis.} $\int_{-1}^{1} x \ofpd dx = F(t)$. Carrying out this 
conversion on \eqref{ndpd1}
\begin{equation} \label{ofpd1}
\ofpd = \frac{1-\xi^2}{(1-x)(x^2 - \xi^2)} \int \frac{d\kperp}{2 (2\pi)^3} \psi \Big( \frac{x+\xi}{1+\xi},\kperp \Big)
 \psi^{*}\Big(\frac{x- \xi}{1- \xi} ,\kperp + (1-\bar{x}) \mathbf{\Delta}^{\perp} \Big),
\end{equation} 
we find our earlier result \cite{Tiburzi:2001ta,comment} for $x>\xi$. 

 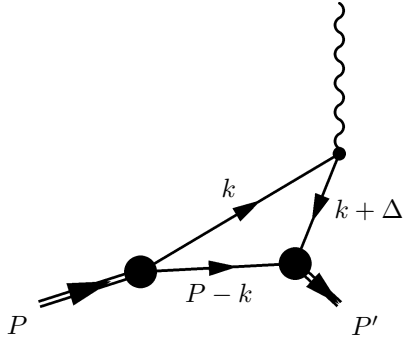
\begin{figure}
 \begin{center}
 \begin{fmffile}{fmfZ}
 \begin{fmfchar*}(40,40)
 	\fmfstraight
   \fmfleft{em,space}  \fmflabel{$P$}{em} 
 	\fmfcurved  
 \fmf{heavy,tension=1.2}{em,Zee}
     \fmf{fermion,tension=.3,label=$k$,label.side=left}{Zee,oh}
     \fmf{photon,tension=1.2}{my,oh} 
     \fmf{fermion,tension=.4,label=$k+\Delta$,label.side=left}{oh,Zam}
     \fmf{fermion,tension=.4,label=$P-k$,label.side=right}{Zee,Zam}
   \fmf{heavy}{Zam,fb}
 	\fmfstraight
   \fmfright{fb,oh,my} \fmflabel{$P^{\prime}$}{fb} 
   	\fmfcurved
 	\fmfdot{oh}
   \fmfv{decor.shape=circle,decor.filled=full,decor.size=.1w}{Zee,Zam}
 \end{fmfchar*}
 \end{fmffile}
 \caption{Time-ordered triangle: the $Z$-graph}
 \label{fZ}
 \end{center}
 \end{figure}

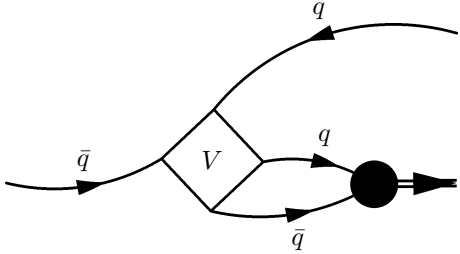
\begin{figure*}
\begin{center}
\begin{fmffile}{fmfZcross}
\begin{fmfchar*}(60,20)
    \fmfstraight
    \fmfleft{em,space}   
    \fmfpolyn{empty,tension=.5,label=$V$}{Z}{4}	
    \fmf{fermion,tension=1,right=.2,label=$\bar{q}$,label.side=left}{em,Z4}
    \fmf{fermion,tension=.3,right=.3,label=$q$,label.side=right}{oh,Z3}
    \fmf{fermion,tension=.3,left=.2,label=$q$,label.side=left}{Z2,Zam}
    \fmf{fermion,tension=.3,right=.2,label=$\bar{q}$,label.side=right}{Z1,Zam}
    \fmf{heavy}{Zam,fb}
    \fmfright{fb,oh} 
    \fmfv{decor.shape=circle,decor.filled=full,decor.size=.1w}{Zam}
\end{fmfchar*}
\end{fmffile}
\vskip1cm
\caption{How the interaction crossing symmetry fixed up the 
non-wave function vertex.}
\label{fZcross}
\end{center}
\end{figure*}

Now when $0 < \kplus < - \Delta^{+}$, we close the contour in the lower-half plane and the integral is 
$- 2 \pi i \Res(k_{b}^{-})$. Once again temporarily ignoring the pre-factors, the residue gives (see Figure \ref{fZ})
a wave function from the initial meson's vertex:
 \begin{equation} \label{qq}
 \Res(k_{b}^{-}) \propto  \psi(X, \kperp + X \Delta^{\perp}/2 ) \bigdw(Y, \kperp + (1-Y/2)\Delta^{\perp}|M^2)
\Gamma^{*}(Y,\kperp + (1-Y/2) \Delta^{\perp}).  \notag
 \end{equation}
 The final meson's vertex, however, represents meson production off the quark line in this time ordering 
 and thus is a non-wave function vertex. Said another way, since $Y = \frac{X - \zeta}{1-\zeta} < 0$, Eq. \eqref{wf} does not 
hold.\footnote{We could define a non-valence wave function from the non-wave function vertex but Eq. \ref{krass} shows how 
this non-valence wave function would be related by crossing to the valence one.} 
 The vertex function, on the other hand, is defined for $Y < 0$ since the physics of a non-wave function vertex is 
 related by crossing symmetry to that of the wave function vertex. Using the Bethe-Salpeter equation \eqref{BS} for the 
 vertex function, we can perform the crossing by inserting the interaction
 \begin{equation} \label{krass}
 \Gamma(Y,\kperp) = \int \frac{dZ d\kaperp}{2 (2\pi)^3 Z(1-Z)} 
 V(Z,\kaperp; Y, \kperp) \psi(Z,\kaperp) ,
 \end{equation}
 which is well defined since only the interaction contains the negative momentum fraction. The interaction has 
 enabled us to cross lines and recover a wave function at the final meson's vertex, see Figure \ref{fZcross}. 
We will comment on dealing with the crossed interaction in light front field theory (using the Wick-Cutkosky model as 
an example) below in section \ref{WFNS}. Having crossed lines, we arrive at
 \begin{multline} \label{ndpd2}
\ndpd =  \frac{-\zeta}{X(1-X)(X-\zeta)} \int \frac{d\kperp d\kaperp dZ}{[2 (2\pi)^3]^2 Z(1-Z)} 
\psi(X,\kperp + X \Delta^{\perp}/2) \\
\times \bigdw \Big(\sigma, \kperp + \sigma \Delta^{\perp} \Big| t \Big) 
V \Big(Z, \kaperp; Y, \kperp + (1-Y/2) \Delta^{\perp} \Big) \psi^{*}(Z,\kaperp), 
\end{multline}  
with $\sigma = X/\zeta$. 

Without interactions at the photon vertex, $\bigdw$ is an approximation to the $q\bar{q}$ component of the photon's wave function.
When $\zeta \to 0$, this approximation does not cause any harm since we recover the Drell-Yan-West
formula (see Appendix B). Additionally, the above expressions guarantee that the SPD is continuous at the crossover ($X = \zeta$). 
This is because for a physically reasonable potential
(one which vanishes at least linearly at the end points), Eq. \eqref{wein} mandates the wave function vanishes quadratically (as pointed
out in \cite{Tiburzi:2000je}). 
Thus Eq. \eqref{ndpd1} vanishes as $X \to \zeta$ due to the wave function and Eq. \eqref{ndpd2} vanishes in the same limit
due to the interaction (in both cases the offending $ X - \zeta$ in the denominator is canceled and what remains vanishes linearly). 
Continuity is required by the factorization theorem for the hard DVCS subprocess. Additionally, due to the reality of the leading-order 
Bethe-Heilter amplitude, $\text{Im}(M^{\mu \nu}) \propto F(\zeta,\zeta,t)$ is physically observable \cite{Diehl:1997bu} 
which is another reason continuity is essential. 

Treating the photon vertex as bare, however, 
requires high $|t|$, which is not only an extraneous assumption but also not experimentally realizable.\footnote{The bare-coupling
vertex is actually of questionable validity even at high $|t|$. This is because the Born approximation to SPD's does not result in the 
bare-coupling SPD's. See section \ref{APPROX}.} Furthermore for the sake of the sum rule, we must make sure the effects of all the 
Fock space components are present in this non-perturbative scheme. To do so, we must include interactions at the photon vertex.

While the bare-coupling SPD's derived above aren't of general applicability, they are simple enough to demonstrate
the required symmetry properties of SPD's, and hence shall not be immediately discarded. To form the SPD for our flavor 
singlet meson in the symmetrical variables, we use \cite{Radyushkin:1997ki}
\begin{equation} \label{convert}
(1 + \xi)^{-1} \ofpd = \ndpd \theta[X(1-X)] + \mathcal{F}(\zeta - X,\zeta,t))\theta[(\zeta - X)(X + 1 - \zeta)],
\end{equation}
with which we can test time reversal invariance which requires invariance under $\xi \to -\xi$. 
Firstly, however, it is instructive to investigate $\ofpd$ for $x < -\xi$ which is 
$\mathcal{F}(\zeta - X, \zeta,t)$ for $\zeta - 1 < X < 0$. Straightforward algebra shows $F(x,\xi,t) = F(|x|,\xi,t)$ for 
$x < - \xi$. Furthermore, the form of equation \eqref{ofpd1} shows $F(x, -\xi, t) = \ofpd$ in accordance with 
time reversal invariance. These observations show we are on the right track. 

In the central region $-\xi < x < \xi$ the $\xi$ symmetry is obscured.  Writing out the consequence
of equation \eqref{convert} in the central region, the $x \to -x$ symmetry is obvious
\begin{multline} 
\ofpd = \frac{- 2 \xi (1 + \xi)}{(x^2 - \xi^2) (1-x)} \int \frac{d\kperp d\kaperp dZ}{4 (2 \pi)^6 Z (1-Z)} 
\psi \Big(X,\kperp + X \Delta^{\perp}/2 \Big) \bigdw \Big(\frac{1}{2} + \frac{x}{2 \xi}, \kperp + \big(\frac{1}{2} + \frac{x}{2 \xi} 
 \big) \Delta^{\perp} \Big| \; t \; \Big)  \\ 
\times V\Big(Z,\kaperp; Y, \kperp + (1 - Y/2) \Delta^{\perp}\Big) \psi^{*}( Z,\kaperp ) + \Bigg\{ x \to -x \Bigg\} \label{ofpd2}
\end{multline}
and follows directly from Eq. \eqref{convert}.
Note that we now use $X$ and $Y$ as replacements for $\frac{x+\xi}{1+\xi}$ and $\frac{x-\xi}{1-\xi}$, respectively. When we separately
calculate the expression for the time-reversed process in Figure \ref{fZ}, we arrive back at the complex conjugate of \eqref{ofpd2}. 
On the other hand, substituting $\xi \to - \xi$ in Eq. \eqref{ofpd2} in order to verify time-reversal invariance is fraught with 
difficulty. Evoking this transformation sends $X \to Y$ and $Y \to X$, and formally the above expression is zero since 
\mbox{$\psi(Y<0,\ldots) =0$}.
	To investigate time-reversal invariance via $\xi$ symmetry, we must do so before appealing to crossing symmetry, since in the
reverse course of events the wave function and non-wave function vertices are switched. Furthermore one must reanalyze the locations of 
the poles, etc. Indeed after much algebraic man{\oe}vering, one recovers Eq. (\ref{ofpd2}) for $\Delta \to -\Delta$ (the details are
relegated to Appendix A). At this point, we have demonstrated the $x$ symmetry as well as the time-reversal invariance
of the bare-coupling SPD's. The $x$ symmetry will obviously extend once we include interactions due to the form of Eq. \ref{convert}.
On the other hand, showing the $\xi$ symmetry will require a slight reworking, though the scheme set up in the appendix is easily 
extended.
 
\section{Interactions at the photon vertex} \label{T}
For general applicability at all values of $t$, 
we saw above that interactions at the photon vertex must be included 
and so we derive the T-matrix for $qq$ scattering by adapting the method in \cite{Callan:1976ps}. 

The T-matrix satisfies a four-dimensional Lippmann-Schwinger equation. We are interested in 
its three dimensional, light-front reduction which gives us the equation
\begin{align} \label{LS}
T(x,\kperp;y,\pperp | R^2) & \equiv \langle y, \pperp; R^2 | \; \hat{T} \; | x, \kperp; R^2 \rangle \notag \\
& = - V(x,\kperp; y,\pperp) + \int \frac{dz d\qperp}{2 (2 \pi)^3 z(1-z)} V(x,\kperp;z,\qperp) 
\bigdw (z, \qperp| R^2) \; T(z, \qperp; y, \pperp | R^2 ),
\end{align}
where $R^2$ is the invariant mass of the system and we have lazily dropped the relative label from all transverse momenta.

Now the function $\phi$ defined by 
\begin{equation} \label{phi}
\phi(x,\kperp;y,\pperp | R^2) = \bigdw (x,\kperp | R^2) \; T(x,\kperp;y,\pperp | R^2)
\end{equation}
satisfies the inhomogeneous equation
\begin{equation}
\Big(R^2 - \hat{H} \Big) \phi(x,\kperp;y,\pperp | R^2) =  - V(x,\kperp; y,\pperp).
\end{equation}
Using the properties of the Green's function (\ref{green}), we can construct the solution of this 
equation, namely
\begin{equation} \label{phieq}
\phi(x,\kperp;y,\pperp | R^2) =  - \int \frac{dz d\qperp}{2 (2 \pi)^3 z(1-z)} G(x,\kperp; z,\qperp | R^2) V(z,\qperp; y,\pperp),
\end{equation}
and in turn write a familiar new expression for $T$
\begin{multline} \label{VGV}
T(x,\kperp;y,\pperp |  R^2 ) =  - V(x,\kperp; y,\pperp)\\
-  \int \frac{dz d\qperp dz^{\prime} d\qperp^{\prime} }{[2 (2 \pi)^3]^2 z(1-z) z^{\prime} (1-z^{\prime})} V(x,\kperp;z,\qperp) 
G(z,\qperp; z^{\prime}, \qperp^{\prime} | R^2) V(z^{\prime}, \qperp^{\prime};y,\pperp).
\end{multline}

Lastly we can render the above expression for $T$ more useful by utilizing the equation of motion for the Green's function
Eq. \eqref{greeneq} to rewrite equation \ref{phieq} as
\begin{equation}
\phi(x,\kperp;y,\pperp | R^2) = 2 (2\pi)^3 \delta(x - y) \delta^2(\kperp - \pperp) - G(x,\kperp;y,\pperp|R^2) \bigdw^{-1}(y,\pperp|R^2).
\end{equation}
Using this information, we arrive at the final form for the T-matrix which will be relevant for quark-quark scattering
at the photon vertex
\begin{multline} \label{bigT}
T(x,\kperp;y,\pperp | R^2) =  \bigdw^{-1}(x,\kperp | R^2)\\
\times \Bigg( 2(2\pi)^3 x (1-x) \delta(x-y) \delta^2(\kperp - \pperp) - G(x,\kperp; y,\pperp | R^2) \bigdw^{-1}(y,\pperp | R^2)  \Bigg). 
\end{multline}

To add the necessary quark-quark interactions, we replace the bare photon coupling $D^{\mu}$ with the photon vertex function 
$\Gamma^{\mu}$, which can be written in terms of the T-matrix: $\Gamma^{\mu} = D^{\mu} + D^{\mu} G_{o} T$. To calculate the 
vertex function, we look at the plus component and perform the light-front reduction. Let the initial quark have momentum
$p$ and the scattered quark $p + \Delta$. Define $W = \pplus/\Pplus$, where $\Pplus$ is some external plus momentum,
and keep $t = \Delta^2$ as above. We thus have
\begin{equation}
\Gamma^{+}(p,-\Delta) / i  \Pplus = (2 W - \zeta) - \zeta \int \frac{dX d\kperp \theta[X(\zeta -X)]}
{2 (2\pi)^3 X(\zeta -X)} (2X - \zeta) \bigdw(\sigma,\kperp | t) T(\sigma,\kperp;\omega,\pperp_{\text{rel}} | t ) , 
\end{equation}
where $\pperp_{\text{rel}} = \pperp + \omega \Delta^{\perp}$, $X = \kplus/\Pplus$ and $\Delta^{+} = - \zeta \Pplus$, 
$\sigma = X/\zeta$ and $\omega = W/\zeta$. Notice that as 
a result of the light-front reduction, $X$ is restricted to $(0,\zeta)$.  
Now we merely insert the T-matrix derived above Eq. \ref{bigT} to discover 
\begin{equation} \label{gamma}
\Gamma^{+}(p,-\Delta) / \Pplus = - i \zeta \bigdw^{-1} (\omega,\pperp_{\text{rel}} | t) \int \frac{dX d\kperp \theta[X(\zeta - X)]} 
{2 (2\pi)^3 X(\zeta-X)} (2X - \zeta) G(\sigma,\kperp; \omega,\pperp_{\text{rel}} | t ). 
\end{equation}
In particular, we should note that the additive bare-coupling term canceled. The above equation tells us how to add $qq$
interactions within the light-front reduced Bethe-Salpeter approach.   

\section{SPD's} \label{AGAIN}
In the non-perturbative Bethe-Salpeter approach to SPD's, we require quark-quark interactions at the photon vertex. Above we
have derived an expression to include them in this approach. Thus we now proceed to derive the SPD's using
everything developed so far.

Following the above, we merely insert the photon vertex function into equation \ref{covartri}, now relabeling $k$ as $p$ since
there are two loop integrals (the $k$ loop integral is contained in $\Gamma^{+}$ and $X \equiv \kplus/\Pplus$ is fixed, 
while $W \equiv \pplus/\Pplus$ is integrated over). There is no new $p^{-}$ dependence introduced and consequently the integration
proceeds exactly as in section \ref{TD} and hence there are two contributions to assess depending on $W$.

For $0<W< \zeta$, we can get the contribution from Eq. \ref{ndpd2} by inserting the form of $\Gamma^{+}$ Eq. \ref{gamma}---being
careful, of course, to remove the constants, derivative coupling, as well as the integral over $X$. This gives a contribution to the SPD 
for $0 < X < \zeta$
\begin{multline} \label{Fb}
\mathcal{F}_{\text{b}}(X,\zeta,t) = - \frac{\zeta^{2}}{X (\zeta - X)} \int \frac{d\kperp dW d\pperp dZ d\kaperp \theta[W(\zeta - W)]}
{[2(2\pi)^3]^3W(1-W)(W-\zeta)Z(1-Z)} \psi(W, \pperp + W \Delta^{\perp} /2)\\ 
\times G(\omega, \pperp + \omega \Delta^{\perp} ; \sigma,\kperp | t ) V \big( Z, \kaperp; Y, \pperp + (1 - Y/2) \Delta^{\perp} \big) 
\psi^{*}(Z, \kaperp). 
\end{multline}
where now we have used $Y = \frac{W - \zeta}{1 - \zeta}$.

On the other hand, appealing to Eq. \ref{ndpd1} for $\zeta < W < 1$ and inserting $\Gamma^{+}$ leads us to $G(\omega > 1, \ldots)$. 
In order to interpret the negative momentum fraction (here $1 - \omega$), we must first use the equation of motion for the 
Green's function (\ref{greeneq}) \cite{alt}:
\begin{multline} \label{greeqom}
\bigdw^{-1} (\omega, \pperp | t) G(\omega, \pperp; \sigma, \kperp  | t) = 2 (2\pi)^3 X (1 - \sigma) \delta(X - W)
\delta^{2} (\kperp - \pperp)\\
+ \int \frac{dZ d\kaperp}{2 (2\pi)^3 Z(1-Z)} V(Z,\kaperp;\omega, \pperp) G(Z,\kaperp; \sigma, \kperp | t ),
\end{multline} 
after which the problematic momentum fraction has been relocated solely in the interaction. As before, the definition of $V$ 
can be extended in our light-front field theory by crossing. In the first term above, the delta function 
mandates $\zeta < X < 1$. While in the second term, we have $0 < X < \zeta$ as originally required by equation \ref{gamma}. Indeed 
the effect of the first term is to regenerate the bare photon coupling, which gives a contribution $\mathcal{F}_{\text{a}}(X,\zeta,t)$ 
identical to Eq. \ref{ndpd1} above \footnote{The simplicity of the SPD for $X>\zeta$ in this approach is related to the expression 
for the form factor. We make this connection explicit in Appendix B}. The second term in Eq. \ref{greeqom} gives rise to the final 
contribution
\begin{multline} \label{Fc}
\mathcal{F}_{\text{c}} (X, \zeta, t) = \frac{\zeta(1-\zeta)}{X(\zeta -X)} \int \frac{d\kperp dW d\pperp dZ d\kaperp \theta[(1-W) 
(W - \zeta)]}{[2 (2\pi)^3]^3 W(1-W)(W-\zeta)Z(1-Z)} \psi(W, \pperp + W \Delta^{\perp}/2) \\
\times V(Z, \kaperp; \omega, \pperp + \omega \Delta^{\perp}) G(Z, \kaperp; \sigma, \kperp | t ) \psi^{*}\big(Y, \pperp + (1-Y/2) 
\Delta^{\perp} \big).
\end{multline}

\begin{figure}
\begin{center}
\begin{fmffile}{fSPD}
	\parbox{40mm}{
	\begin{fmfgraph}(40,30)
	\fmfleft{a,b,c}
	\fmf{double,tension=1}{a,in}
	
	\fmfpolyn{shade,tension=.3}{Z}{4}
	\fmf{plain,tension=.5,right=.2}{in,Z4}
        \fmf{plain,tension=.2,right=.2}{Z3,Zee}
	\fmf{plain,tension=.45,left=.1}{Z2,Zam}
    	\fmf{plain,tension=.35,right=.1}{Z1,Zam}
 	
	\fmf{plain,tension=.2}{in,Zee}
	\fmf{double,tension=.8}{Zee,out}
    	\fmf{double}{Zam,fa}
    	\fmf{plain}{out,fc}
	\fmf{plain}{out,fd}
	\fmfright{fa,space,fc,fd} 
    	
	\fmfv{decor.shape=circle,decor.filled=full,decor.size=.07w}{in,out,Zee,Zam}
    	
	\end{fmfgraph}} +
	\parbox{40mm}{  	
	\begin{fmfgraph}(40,30)
	\fmfleft{a,b,c}
	\fmf{double,tension=1}{a,in}
  	
	\fmfpolyn{shade,tension=.3}{Z}{4}
	\fmf{plain,tension=.3,left=.2}{in,Z3}
        \fmf{plain,tension=.2,left=.2}{Z4,Zam}
	\fmf{plain,tension=.5,left=.1}{Z2,Zee}
    	\fmf{plain,tension=.3,right=.1}{Z1,Zee}
 	
	\fmf{plain,tension=.1}{in,Zam}
	\fmf{double,tension=.8}{Zee,out}
    	\fmf{double}{Zam,fa}
    	\fmf{plain}{out,fc}
	\fmf{plain}{out,fd}
	\fmfright{fa,space,fc,fd} 
    	
	\fmfv{decor.shape=circle,decor.filled=full,decor.size=.07w}{in,out,Zee,Zam}
    	
    	\end{fmfgraph}}
\end{fmffile}
\end{center}
\caption{Graphical representation of contributions to the SPD for $X < \zeta$: $\mathcal{F}_{b} + \mathcal{F}_{c}$. The shaded box represents the crossed interaction $V(1 \to 3)$, and we implicitly include a sum over all meson states for the internal double line (thus effectively
the double line represents the four-point Green's function).}
\label{SPDfig}
\end{figure}
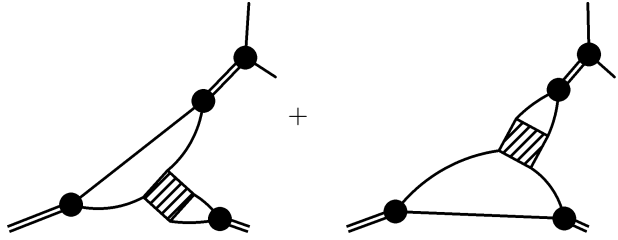

Combining these results, we can finally write the skewed parton distribution as
\begin{equation} \label{it}
\ndpd = \mathcal{F}_{\text{a}}(X, \zeta,t) \theta[(X - \zeta)(1-X)] + \Big( \mathcal{F}_{\text{b}}(X, \zeta,t) 
+ \mathcal{F}_{\text{c}}(X, \zeta, t) \Big) \theta[X(\zeta - X)],
\end{equation}
which is the $3+1$-dimensional generalization of the SPD's found in \cite{Burkardt:2000uu}.
We should note: given the form of the Green's function (\ref{green}), the continuity of the SPD above is maintained at $X = \zeta$ due, 
this time, solely to the nature of the wave function at the end points. 
The graphical representation of the SPD for $X<\zeta$ is shown in Figure \ref{SPDfig}.

\section{Approximating the Green's function} \label{APPROX} 
With Eq. \ref{it}, we have deduced the form of the SPD satisfying the necessary symmetry properties as well as the continuity
condition at the crossover. It remains to check whether the sum rule, Eq. \ref{sum} is satisfied, however, evaluating an eight-dimensional
integral for the SPD (or nine-dimensional for the sum rule) is not exactly the best way to proceed. In this section, we investigate
possible approximations for the Green's function which are still consistent with the general properties of the SPD. Of course, there
is the obvious simplification if $t$ is near a bound state pole. But in general, one seeks to avoid solving for the entire spectrum 
as well as the eigenfunctions of a given light-front potential. Additionally, there is the Born approximation for high $|t|$ which 
for $\mathcal{F}_{b}$ (Eq. \ref{Fb}) results in the bare-coupling SPD encountered above. There is also the contribution  
from $\mathcal{F}_{c}$ (Eq. \ref{Fc}) at high $|t|$ which is absent from the bare-coupling SPD.   

\subsection{Non-relativistic approximation} \label{nrap}
In the non-relativistic approximation, the integral equation for the T-matrix becomes algebraic. Thus one can obtain 
an approximation of the Green's function by way of the T-matrix. Let us return to Eq. \eqref{LS} for the T-matrix. 
In a non-relativistic scenario ($m \to \infty$), we may write $\bigdw \to \dw$ assuming the self interactions become
negligible in this limit, i.e. $m^2 f(m^2) \to 0$. And thus
\begin{align}
T(x,\kperp;y,\pperp | R^2)  + V(x,\kperp;y,\pperp)  & \approx  V(x,\kperp;1/2,0) \int \frac{dz d\qperp}{2 (2 \pi)^3 z(1-z)} 
\dw(z,\qperp | R^2) T(z,\qperp;y,\pperp | R^2 ) \notag \\
    & \equiv  - V(x,\kperp;1/2,0) I(y,\pperp | R^2).
\end{align}
The resulting unknown integral $I(y,\pperp | R^2)$ can be solved for algebraically. The result leaves us with the above approximation 
to the T-matrix with
\begin{equation} \label{I}
I(y,\pperp | R^2) = \frac{1}{1 - C(m^2,R^2)} \int \frac{dz d\qperp}{2 (2\pi)^3 z(1-z)} \dw(z,\qperp | R^2) V(z,\qperp;y,\pperp),
\end{equation}
where
\begin{equation}
C(m^2,R^2) = \int \frac{dz d\qperp}{2 (2\pi)^3 z(1-z)} \dw(z,\qperp|R^2) V(z,\qperp;1/2,0).
\end{equation}
Further simplification of Eq. \ref{I} in the non-relativistic limit is not possible, since doing so results in 
a logarithmically divergent expression. 

Inserting the approximate T-matrix into equation \ref{bigT}, we can in turn produce the non-relativistic
approximation to the Green's function
\begin{equation} \label{nrgreen}
G^{\text{NR}}(x,\kperp; y,\pperp|R^2) \approx 4 \pi^3 \delta(y - 1/2) \delta^2(\pperp) I(x,\kperp |  R^2) \dw(x ,\kperp | R^2), 
\end{equation} 
where is it understood that this expression is to be used where the variables $y$ and $\pperp$ are integrated over.
Notice that as $x \to 1$, the propagator vanishes which is only enough to yield a finite SPD at the crossover. The additional
term $I(x,\kperp|R^2)$ depends upon the interaction. Using general properties of $V$ from the Weinberg equation
and electromagnetic form factor, one can show $I(x = 1,\ldots) = 0$ \cite{Tiburzi:2000je}. (This is just a way to motivate the physicality
of requiring $V$ to vanish linearly at the end points.) This feature applied to SPD's 
maintains continuity at the crossover between valence and non-valence r\'egimes ($X = \zeta$).

In the non-relativistic approximation scheme (NR), we arrive at the distributions
\begin{equation}  \label{NRNV}
\mathcal{F}_{\text{b}}^{\text{NR}} + \mathcal{F}_{\text{c}}^{\text{NR}}  = \Big( C_{\text{b}}(\zeta,t) + C_{\text{c}}(\zeta,t) \Big)  
\int \frac{d\kperp}{2 (2\pi)^3 X (\zeta - X)} I(\sigma, \kperp | t) \dw(\sigma, \kperp | t )
\end{equation}
where we have defined
\begin{align}  \label{NRCs}
C_{\text{b}}(\zeta,t) & =  \frac{\zeta}{1 - \zeta/2} \psi \Big( \zeta/2, \frac{\Delta^{\perp}}{2} (\zeta/2 -1)\Big)
V \Big(\frac{1}{2},\mathbf{0}^{\perp};-\frac{\zeta/2}{1-\zeta}, \frac{\mathbf{\Delta}^{\perp}}{2}\big( 1 + \frac{\zeta/2}{1-\zeta}\big)\Big)
\psi^{*}(\mathbf{r} = 0) \notag \\
C_{\text{c}}(\zeta,t) & =  \frac{\zeta (1-\zeta)}{2(2\pi)^3} \int \frac{dW d\pperp \theta[(W-\zeta)(1-W)]}{W(1-W)(W - \zeta)} \psi(W, \pperp + W \Delta^{\perp}/2) 
V(1/2, \mathbf{0}; \omega, \pperp+ \omega \Delta^{\perp}) \psi^{*} \big(Y, \pperp + (1- Y/2) \Delta^{\perp} \big) ,
\end{align}
and we have suggestively written
\begin{equation} \label{origin}
\psi(\mathbf{r}=0) = \int \frac{dz d\kperp}{2(2\pi)^3 z(1-z)} \psi(z,\kperp).
\end{equation}

\subsection{Closure approximation} \label{clap}
The non-relativistic approximation to the Green's function is an important limiting case but not of general applicability. 
Here we use the closure approximation which can be applied to the SPD of any system by calculating two parameters. 
Into the energy denominator of the Green's function, let us introduce the parameter $\mn$
\begin{align} \label{expand}
\frac{1}{R^2 - M^{2}_{n}} & = \frac{1}{R^2 - \mn + \mn - M^{2}_{n}} \notag \\
	& =  \frac{1}{R^2 - \mn} - \frac{\mn - M^{2}_{n}}{(R^2 - \mn)^2} + \ldots
\end{align}
Now requiring the contribution to the SPD from the second term in the Green's function expansion 
to vanish puts a restriction on $\mn$. Of course there are really two parameters $\mn_{b}$ and $\mn_{c}$ which could 
be calculated from having the resulting integrals in Eqs. (\ref{Fb}, \ref{Fc}) vanish. Exact calculation of the
parameters, however, requires the exact Hamiltonian. But in principle, the expansion is well defined.

Keeping only the first term in equation \ref{expand} and using the completeness relation (\ref{compl}) we arrive at the
distributions in the closure approximation (CL)
\begin{align} \label{CLNV}
\mathcal{F}_{\text{b}}^{\text{CL}} & = \frac{\zeta}{t - \mn_{b}} \int \frac{d\kperp dZ d\kaperp}{[2(2\pi)^3]^2 X(1-X)(\zeta - X) Z(1-Z)} 
\psi(X, \kperp) V \big( Z, \kaperp; Y, \kperp - (1 - \bar{x}) \Delta^{\perp} \big) \psi^{*}(Z, \kaperp)   \notag \\ 
\mathcal{F}_{\text{c}}^{\text{CL}} & = \frac{\zeta (1-\zeta)}{t - \mn_{c}} \int \frac{d\kperp dW d\pperp \theta[(1-W)(W - \zeta)]}
{[2(2\pi)^3]^2 W(1-W)(W - \zeta)X(\zeta - X)} \psi(W, \pperp + W \Delta^{\perp}/2) V(\sigma,\kperp;\omega,\pperp_{\text{rel}})
\psi^{*}(y,\pperp + y \Delta^{\perp} /2)
\end{align}
where above $Y = \frac{X - \zeta}{1 - \zeta}$, $\bar{x} = (X + Y)/2$ and $y = \frac{W - \zeta}{1 - \zeta}$. The finite-ness of the 
above expressions depends solely on the limiting behavior of the crossed potentials. Given a linearly vanishing potential at the
end points, the closure approximated SPD is finite at the crossover, but not zero and hence the SPD is discontinuous. 
This is not surprising 
since the closure approximation to the Green's function is highly similar to a Weinberg propagator but without the necessary
end-point behavior (which forced the bare-coupling SPD's to vanish). It would be possible in a given theory to use the end-point
behavior of the crossed interaction along with the continuity condition to derive a constraint between $\mn_{b}$ and $\mn_{c}$. 
Consequently the sum rule could then be exploited to evaluate the remaining parameter. This will not be pursued here, but will
likely be the source of continued investigation.

\section{Wick-Cutkosky model wave functions} \label{WFNS}
One cannot hope to use wave functions which are not derived from field theory to verify the sum rule in this approach 
since we would be at a 
loss to properly extend the potential's definition. Thus model wave functions derived from \emph{ad hoc} potentials 
(such as those found in our earlier work or the constituent quark model wave functions) cease to be useful. 
Here, we use the Wick-Cutkosky model and begin with a weak binding 
solution in the non-relativistic limit \cite{Karmanov:1980if}. 
Certainly this choice is motivated didactically since it results in 
clean analytic expressions with wave functions about which we have great intuition. Indeed the point of this paper is to demonstrate 
\emph{proof of concept} (via time-reversal invariance and the sum rule) rather than trying to exploit features of SPD's to investigate 
model properties as was attempted in \cite{Choi:2001fc}. 

Applying the Wick-Cutkosky model to our problem is simple, we merely use an interaction proportional to $\sqrt{\alpha} q^2 \phi$ in our 
scalar quark's Lagrangian. Here, $\phi$ is a massless scalar field mediating the interaction. For a suitably weak coupling $\alpha$,
the potential can be well approximated by single $\phi$ exchange. Summing the two light-front, time-ordered, 
one boson exchange diagrams (seen in figure \ref{fOBEP}) gives rise to our potential. Let the total momentum of the system
be $P$ and define $z = \pplus/\Pplus$ and $y = \kplus /\Pplus$. Using the time-ordered rules
(see e.g. \cite{Cooke:2000ef}), we arrive at

\begin{figure}
\begin{center}
\begin{fmffile}{fmfOBEP}
	\parbox{40mm}{
	\begin{fmfchar*}(40,15)
	\fmfleft{a,b}
	\fmf{fermion,tension=.2,label=$p$,label.side=left}{b,in}
	\fmf{fermion}{a,inn}
 	\fmf{gluon}{in,inn}
	\fmf{fermion,label=$k$,label.side=left}{in,d}
	\fmf{fermion,tension=.2}{inn,c}	
	\fmfright{c,d} 
    	\end{fmfchar*}} +
	\parbox{40mm}{  	
	\begin{fmfchar*}(40,15)	
	\fmfleft{a,b}
	\fmf{fermion,label=$p$,label.side=left}{b,in}
	\fmf{fermion,tension=.2}{a,inn}
 	\fmf{gluon}{in,inn}
	\fmf{fermion,tension=.2,label=$k$,label.side=left}{in,d}
	\fmf{fermion}{inn,c}	
	\fmfright{c,d} 
	\end{fmfchar*}}
\end{fmffile}
\end{center}
\caption{Diagrammatic representation of the one boson exchange potential: $V = V_{a} + V_{b}$.}
\label{fOBEP}
\end{figure}
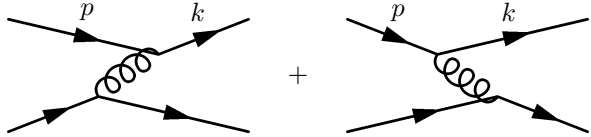
\begin{align} \label{Va}	
V_{a} &= g^2 \frac{\theta(z-y)}{2(\pplus -\kplus)} \frac{1}{P^{-} - p^{-} - (p-k)^{-} - (P-k)^{-}}   \\\label{Vb} 	
V_{b} &= g^2 \frac{\theta(y-z)}{2(\kplus -\pplus)} \frac{1}{P^{-} - k^{-} - (k-p)^{-} - (P-p)^{-}}  ,
\end{align}
where $V_{a}$ corresponds to the first diagram in figure \ref{fOBEP} and $V_{b}$ the second. Here $g^2 = 16 \pi m^2 \alpha$. 
The sum of these two terms can be combined into
\begin{equation} \label{OBEP}
V(z, \kperp; y, \pperp) = - g^2 / \mathcal{K}^2,
\end{equation} 
with
\begin{equation}
\mathcal{K}^2 = \kperp^2 \frac{S(z,y)}{z(1-z)} - 2 \kperp \cdot \pperp + \pperp^2 \frac{S(y,z)}{y(1-y)} + m^2 L(z,y) 
\end{equation}
where $M^2$ is the total invariant mass squared, and we have again lazily dropped the relative label from all transverse momenta. 
Additionally we have defined the replacements
\begin{align}
L(z,y) & = \frac{(z-1/2)(z - y)}{z(1-z)} + \frac{(y -1/2)(y-z)}{y(1-y)} - \frac{M^2}{m^2} |z-y| \notag \\
S(z,y) & = \begin{cases}
		\; z(1-y), \quad  z>y\\
	        \; y(1-z), \quad  y>z
	    \end{cases} \notag 
\end{align}
The wave function is then deduced by solving the Weinberg equation (\ref{wein}) for the one boson exchange potential. In a weakly bound
system, we can neglect to leading order the $M^2$ dependence of this potential by using $M^2 \approx 4 m^2$, and thus the expressions
derived above for skewed parton distributions are valid. 

\begin{figure}
\begin{center}
\begin{fmffile}{fmfCPP}
\begin{fmfchar*}(60,30)
	\fmfstraight
	\fmfleft{yeah,space,spc}
	\fmf{fermion,label=$P-p$,label.side=right}{yeah,wow}
	\fmf{fermion,tension=.2,label=$P - k$,label.side=right}{wow,woho}
	\fmf{gluon,tension=.6}{wow,wee}
	\fmf{fermion,tension=.1,label=$k$,label.side=right}{wee,woo}
	\fmfright{woho,woo,ohya}
	\fmf{scalar,tension=.4,label=$p$,label.side=right}{ohya,wee}
\end{fmfchar*}
\end{fmffile}
\vskip1cm
\caption{Diagram of the crossed-pair potential (hadronization off the quark line) using one boson 
exchange in the Wick-Cutkosky model. The quark line at the top is altered to denote  
traveling backwards in time.}
\label{fCPP}
\end{center}
\end{figure}
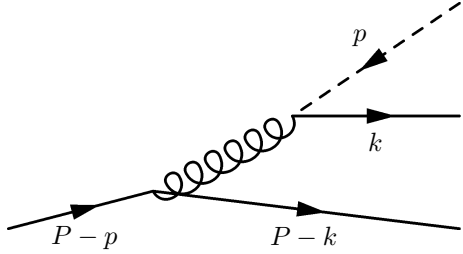

Next we must extend the definition of the potential by crossing quark lines. First we shall consider the case of one negative
momentum fraction as required by the SPD.  Within the Wick-Cutkosky model, we represent the interaction blob in Figure 
\ref{fZcross} by the exchange of one massless boson. We thus use the light-front time ordered graph to calculate the potential 
for quark pair production off the initial quark line.
The interaction which describes hadronization off the quark line is then described by the graph in Figure 
\ref{fCPP}.\footnote{Notice that we do not include a 
spectator quark line of momentum $q$, say. This is because the time-ordering of the Z-graph is fixed by the $q^{-}$ integration. 
Indeed we are merely interpreting the non-wave function vertex. Were we to include an intermediate state with energy $q^{-}$ in 
the denominator of Eq. \ref{Va}, the crossed potential would have the wrong functional dependence.} Notice we omit the diagram
with pair creation from the vacuum. The quark with momentum $p$, has $y<0$ and thus travels backwards in time (as the diagram indicates).
Thus using the time-ordered rules, we arrive at Eq. \ref{Va} for the crossed potential. Furthermore, we can use 
Eq. \ref{OBEP} for the crossed potential since the contribution from $V_{b}$ (which would correspond to pair creation from the
vacuum) vanishes. 

Next we need the potential for one momentum fraction greater than one. If the quark with $z = \pplus/\Pplus >1$, 
emits a $\phi$ boson which produces a quark pair, then one of that pair has the negative momentum fraction $1-z$ and hence
travels backwards through time. Using the time-ordered rules for this process yields an expression identical to $V_{b}$ 
(where now $V_{a}$ vanishes due to vacuum pair production). Thus again we can use Eq. \ref{OBEP} this time for for $z>1$.  
It should not be surprising that the light front potential automatically contains such processes with ``partons 
moving backwards in time'' since the theta functions present in Eqs. (\ref{Va}, \ref{Vb}) do not restrict the momentum fractions
to be between zero and one.

Having the necessary potentials at hand, 
we now attempt to find an approximate solution to the Weinberg equation \eqref{wein} for the one boson, exchange potential
\eqref{OBEP} using the non-relativistic scheme. This method was first used by Karmanov 
\cite{Karmanov:1980if} to determine an approximate solution to the Wick-Cutkosky model, and so we summarize the procedure here.
Following the approximation used in section \ref{nrap}, we can reduce the Weinberg equation to
\begin{equation} \label{nrwein}
\Bigg[\frac{\kperp^2 + m^2}{x(1-x)} - M^2 \Bigg] \psi(x, \kperp) \approx - V \big( \frac{1}{2}, \mathbf{0}; x, \kperp) \int
\frac{dz d\kaperp}{2 (2 \pi)^3 z (1-z)} \psi(z, \kaperp)
\end{equation}
in the limit $m \to \infty$.
The eigenvalue $M^2$ can then be solved for algebraically in the non-relativistic limit. We find $M = 2 m - \frac{1}{4} m \alpha^2$. 
We then deduce the ground state wave function in this approximation by rewriting Eq. \eqref{nrwein}
\begin{equation} \label{wvfn}
\psi(x,\kperp) = \frac{m^3 \sqrt{N} x^2 (1-x)^2}{(\kperp^2 + (2 x - 1)^2 m^2 + x(1-x) m^2 \alpha^2)^2} \frac{1}{1 + |2x -1|} ,
\end{equation}
where $N$ is the normalization constant specified by Eq. \eqref{ortho}. 

\begin{figure}
\begin{center}
\epsfig{file=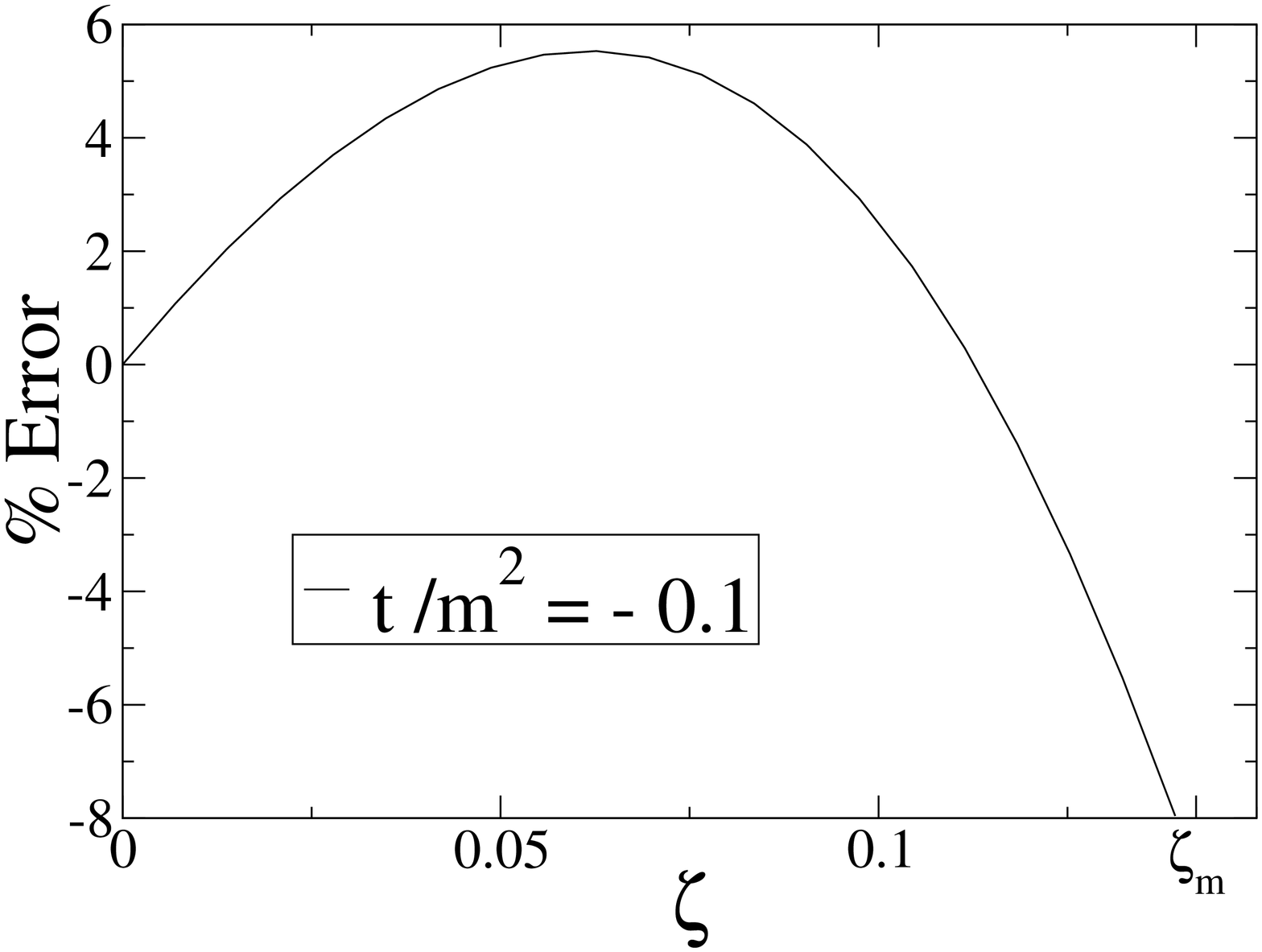,height=2.0in,width=2.0in,angle=270}
\epsfig{file=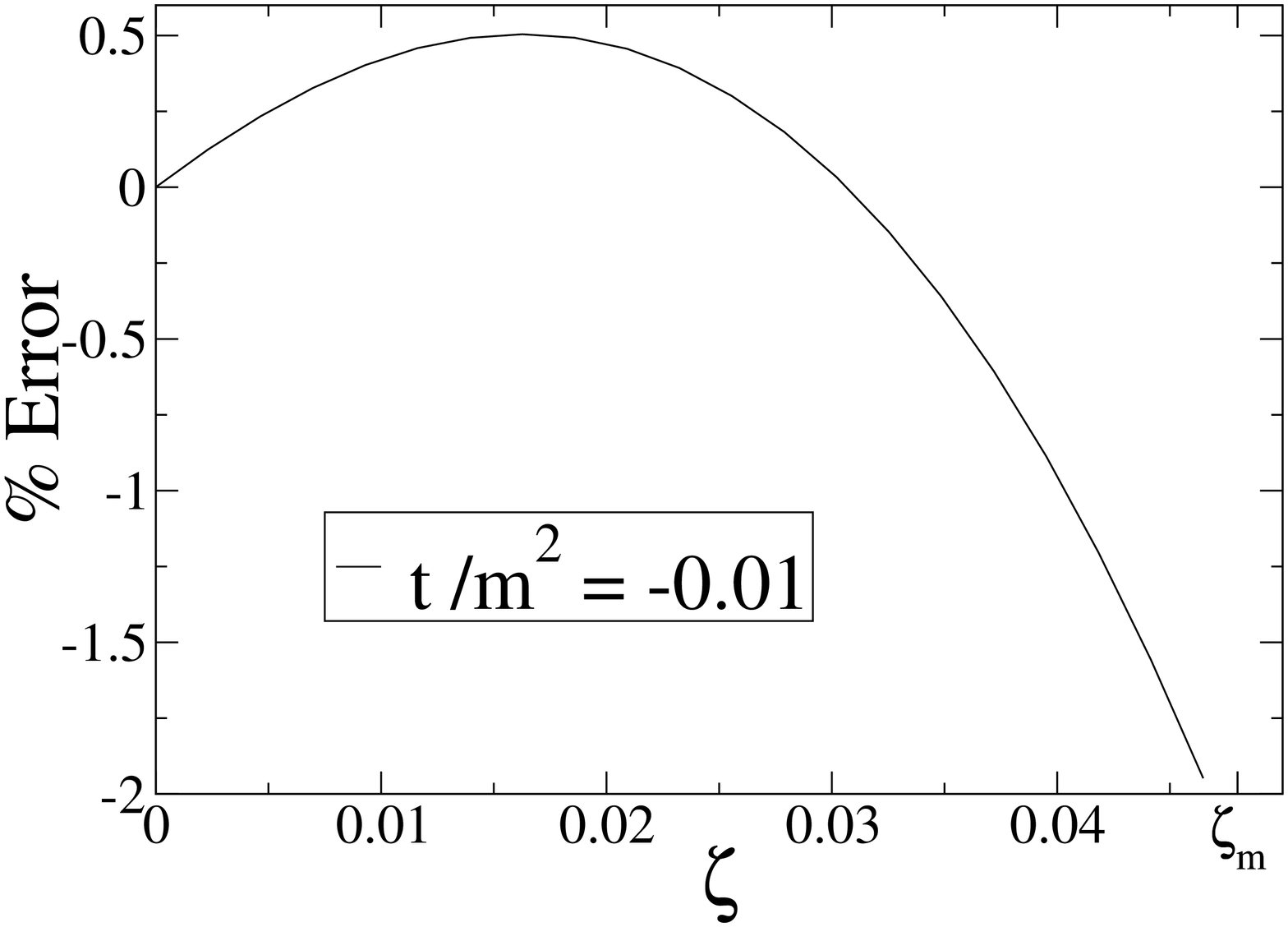,height=2.0in,width=2.0in,angle=270}
\epsfig{file=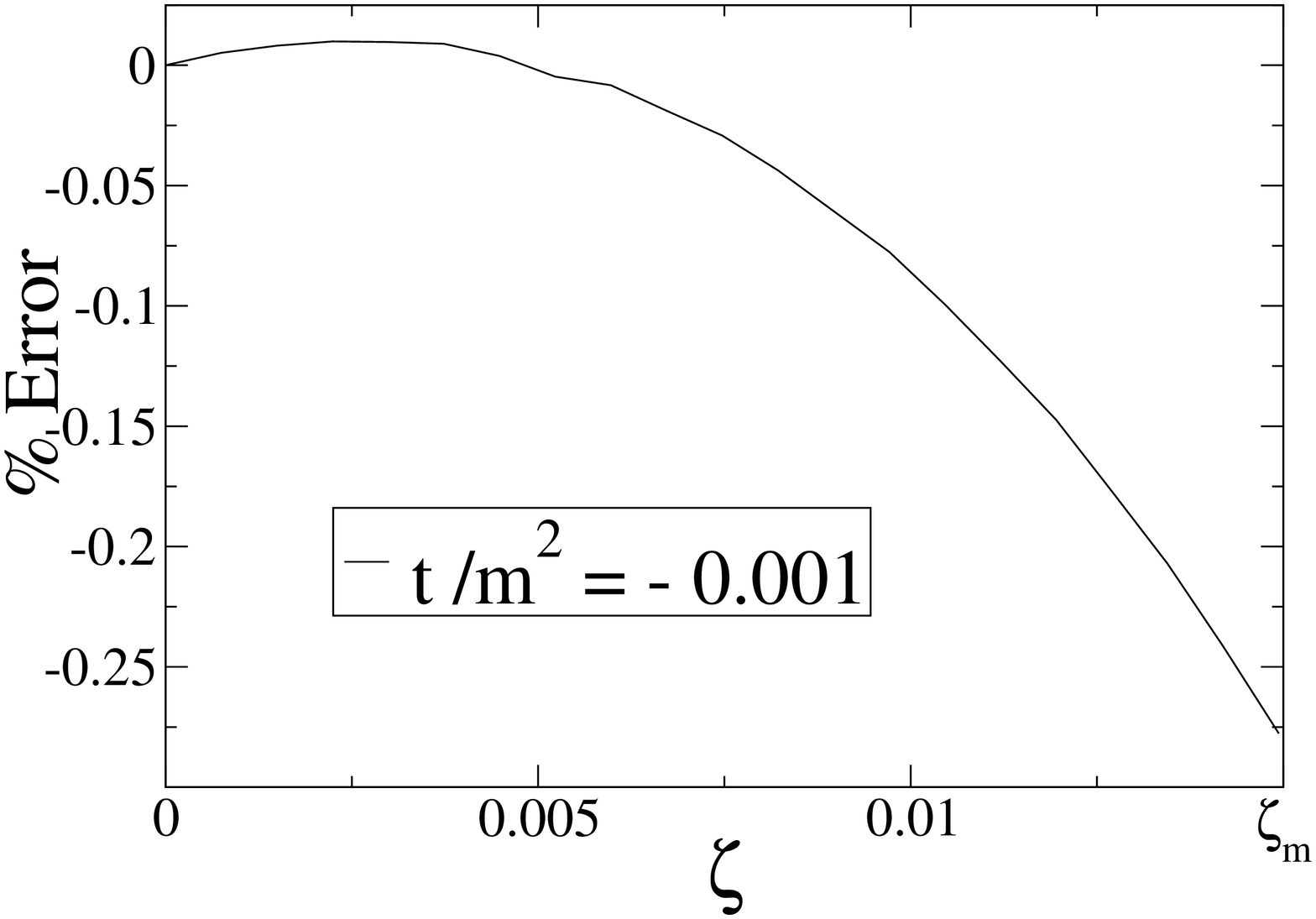,height=2.0in,width=2.0in,angle=270}
\caption{Verification of the sum rule in the non-relativistic limit. The percent error in the determination of the form
factor (from the valence region SPD) plotted as a function of $\zeta$ for small values of $-t/m^2$.}
\label{sumrule}
\end{center}
\end{figure}

This solution and the ensuing non-relativistic
approximation scheme are valid to $\mathcal{O}(\alpha \ln \alpha)$ from the weak binding limit and $\mathcal{O}(-t/m^2)$
from the non-relativistic approximation. When we look at the approximate non-valence SPD's in this scheme
(Eq. \ref{NRNV}), they are at $\mathcal{O}(\alpha^2 \ln \alpha)$ relative to the valence region---the logarithmic
dependence appears since the integral for $C_{c}(\zeta,t)$ in Eq. \ref{NRCs} diverges when $\alpha \to 0$.
This is just as we might suspect in 
a weakly bound theory---pair creation and subsequent interaction are suppressed. Nonetheless, Eq. \ref{NRNV} does give 
us a sense of the non-valence distributions, however, the sum rule Eq. \ref{sum} can be verified to leading order using 
only the valence contribution to the SPD (Eq. \ref{ndpd1}). In Figure \ref{sumrule}, we plot the percent contribution
to the form factor from the valence SPD as a function of the skewness $\zeta$. Lorentz invariance demands 
$\zeta$ independence but our non-relativistic approximation retains $\zeta$ dependence. The sum rule is satisfied by our 
non-relativistic approximation since the corrections are of $\mathcal{O}(-t/m^2)$.

\begin{figure}
\begin{center} 
\epsfig{file=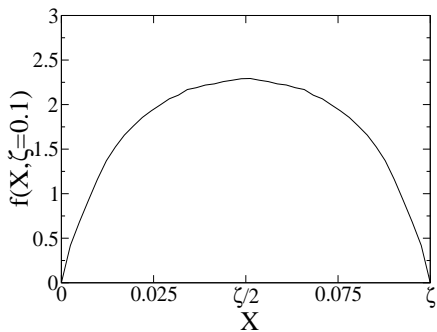,height=2.5in,width=2.0in,angle=270}
\caption{Structure of the non-valence SPD in the non-relativistic limit $-t/m^2 \ll 1$. Here we plot $f(X,\zeta)$ (which is
proportional to $\int I \dw$ appearing in Eq. \eqref{NRNV}) for the fixed value $\zeta = 0.1$.}
\label{nvspd} 
\end{center}
\end{figure}

Despite being suppressed by powers of $\alpha$ relative to the valence region, we can still look at the SPD's structure 
in the non-valence region (Eq. \ref{NRNV}). The coefficient functions $C_{b}(\zeta,t)$ and $C_{c}(\zeta,t)$ can both be calculated
for small $\zeta$ and $-t/m^2$: we find $C_{b} \sim + N \zeta^4 /8$ and 
$C_{c} \sim - N \zeta^2 \times 10^3$,
where $N = 0.00545$ is the normalization constant appearing in Eq. \ref{wvfn} (determined with $\alpha = 0.08$).   
Thus apart from variation with $\zeta$ we can ascertain the non-valence SPD's behavior from 
$\int I \dw$. In Figure \ref{nvspd} we plot $f(X,\zeta) \equiv -\frac{1}{\alpha} \int \frac{d\kperp}{\sigma(1-\sigma)} 
I(\sigma,\kperp|t)\dw(\sigma,\kperp|t)$ as a function of $X$ for fixed $\zeta$. 
The choice of sign preserves the overall sign of $\mathcal{F}$. The figure 
shows the SPD is symmetrical about $X = \zeta/2$ arising from the symmetry of $I$ and $\dw$ about $\sigma = 1/2$ 
(which in turn is due to the fact that our quarks are equally massive). Of course, as a function of $\sigma \in (0,1)$
the profile in the figure is universal. Given the nature of $C_{b}$ and $C_{c}$, since the scaled function
$f(X,\zeta)$ is of order unity, the non-valence SPD is
$\sim 10^6$ times smaller than the valence SPD 
for small $\zeta$ and $-t/m^2$ (here we use $\zeta = 0.1$ for the comparison). 
The additional suppression (beyond powers of the weak coupling) is due to 
accessing the highly concentrated wave function near the end points.
 
Using the non-relativistic limit to verify the sum rule is surely a good start, however, it does away with many of the
complications inherent to the light-front approach. Were we to use the full solution to the Wick-Cutkosky 
model, not only would we have to worry about a consistent approximation scheme for the Green's function, but also
the lack of covariance (here rotational invariance). The above expression Eq. \eqref{it} is exact 
and consequently covariant (although not
manifestly so), but using any realistic procedure to solve for the wave function, we must truncate the interaction kernel. Doing so we
loose covariance, which is approximately restored only when our approximations to the kernel are improved 
\cite{Cooke:2000yi}. This loss of rotational invariance would not only lead to improperly non-degenerate spectra but also 
would tamper with the sum rule. Without manifest Lorentz covariance, Eq. \eqref{sum} will only be approximately satisfied
and a true challenge to verify exactly. This is a difficulty which plagues any Hamiltonian theory.

\section{Conclusion and outlook} \label{conc}
Above we have seen how to express the pion's skewed parton distribution in terms of two body, light-front wave functions. 
By using a light-front reduced Bethe-Salpeter approach, we can express the SPD in terms of two body wave functions
by crossing the interaction (and making the assumption that the effective $qq$ potential is independent of energy). 
We did so first using only bare-coupling at the photon vertex. The functional form derived 
Eqs. (\ref{ndpd1},\ref{ndpd2}) is time reversal invariant (as we explained how to demonstrate through $\xi$ symmetry) 
and is continuous at the crossover. In this non-perturbative scheme, however, the bare-coupling photon vertex is not
justified and thus we derived the photon vertex function Eq. \ref{gamma} in order to include the necessary $qq$ interactions
and hence the effect of higher Fock states. 
The interacting theory SPD (\ref{it}) can then be written in terms of two body wave functions and the Green's function. The form
derived maintains the necessary symmetry properties and is continuous. Potential schemes for approximating the Green's function
were then investigated. We found the Born approximation to the photon vertex does not result in the bare-coupling SPD since
the diagram corresponding to $\mathcal{F}_{c}$ Eq. \eqref{Fc} is neglected. The non-relativistic approximation \eqref{NRNV} and the closure
approximation \eqref{CLNV} to the non-valence region were then derived.  

Using the non-relativistic approximation, we were able to verify the sum rule for the Wick-Cutkosky model in the weak binding 
limit (see Figure \ref{sumrule}). Utilizing the crossed-pair potential and the non-relativistic approximation scheme, 
we explored SPD's in the non-valence region (see Figure \ref{nvspd}). This approach indeed provides an alternative 
to the explicit $N$-body Fock space expansion, and hence a different interpretation of the contributions to the valence and non-valence 
regions. We believe this paradigm is the most useful intuitively, though currently suffers from calculational intractability. 
For instance: the mass squared dependence of the interaction has been neglected here (as is often the case elsewhere, e.g. constituent
quark models) and remains a crucial issue 
to be dealt with, not only in the context of these distributions but for form factors and quark distribution functions as well. Additionally
calculation of the full four-point Green's function presents another technical challenge.
The closure approximation may however provide a way to reasonably approximate the Green's function, since continuity and the 
sum rule can be used as constraints to check how well the full Hamiltonian can be described by the two-body sector. 
Lastly, we find that use of the crossed-pair potential in tandem with the Green's function for the interacting theory allows 
one to obtain non-wave function vertices present in SPD's and time-like form factors.

\begin{center}
{\bf Acknowledgment}
\end{center}
We thank J. R. Cooke for insight about one-boson exchange potentials and the Wick-Cutkosky model.
This work was funded by the U.~S.~Department of Energy, grant: DE-FG$03-97$ER$41014$.  

\section*{Appendix A: $\xi$ symmetry} 
At best, demonstrating the $\xi$ symmetry of our SPD's is subtle. To keep matters simple, we consider only the bare-coupling
SPD's of section \ref{TD} and neglect quark self interactions, since their inclusion is straightforward and tangential
to time-reversal invariance. 
While Eq. \ref{ofpd1} is clearly even in $\xi$, the sign of $\Delta^{+}$ has been reversed and thus the 
location of the poles changed. Thus the whole derivation
isn't necessarily valid for $\xi \to -\xi$.  Furthermore, directly 
switching $\xi \to -\xi$ in equation \ref{ofpd2} yields zero. The time-reversal properties of our results
are a crucial guide to consistency. Separately calculating the time-reversed diagrams of Figures \ref{tri} and \ref{fZ}, we easily 
arrive at the complex conjugates of Eqs. \ref{ofpd1}, \ref{ofpd2}. In order to show the relation between $\xi$-symmetry and
time-reversal, we recalculate the covariant triangle diagram with $\Delta \to -\Delta$. 

Looking at Figure \ref{tri} with $\Delta$ replaced with $-\Delta$,
we have the following expression for the covariant diagram
\begin{equation} \label{tri2}
 -i \int \frac{d^4k}{(2\pi)^4} \frac{(2\kplus- \Delta^{+}) 
\Gamma(k,P) \Gamma^{*}(k-\Delta, P-\Delta)}{(k^2 - m^2 + \ie)
((k-\Delta)^2 - m^2 + \ie)((P-k)^2-m^2 + \ie)}.
\end{equation}
Having reversed the sign of $\Delta^{+}$ in the diagram, we can 
again choose $\Delta^{+} \equiv -2 \xi \Pbarplus \leq 0$, 
where $\xi >0$ and thus we have effectively taken $\xi$ here to be
$-\xi$ of our previous calculation. Due to the kinematics, however,
$\Pplus = (1 - \xi) \Pbarplus$. Thus defining $\Delta^{+} = - 
\Omega \Pplus$ yields $\Omega = \frac{2 \xi}{1 - \xi}$. This should
be compared to the skewness $\zeta = \frac{2 \xi}{1+ \xi}$. We still choose to work in the symmetrical frame, however, this constraint now
reads $\Pperp = \Delta^{\perp}/2$ which is natural given the relative sign difference. 

The poles of the $k^{-}$ integrand are situated at
\begin{equation}
	\begin{cases}
k_{a}^{-} = \Delta^{-} + \frac{m^2 + (\kperp - \Delta^{\perp})^2}{2(\kplus - \Delta^{+})} - \frac{\ie}{2(\kplus
- \Delta^{+})}\notag\\
k_{b}^{-} = \frac{m^2 + \kperp{}^2}{2 \kplus} - \frac{\ie}{2 \kplus}\notag\\
k_{c}^{-} = P^{-} - \frac{m^2 + (\Pperp -\kperp)^2}{2(\Pplus - \kplus)}+\frac{\ie}{2(\Pplus - \kplus)}.\notag
	\end{cases}
\end{equation}
When $\kplus < \Delta^{+}$, all poles lie in the upper-half
complex plane and when $\kplus > \Pplus$, all poles lie in the lower-half plane. Thus, by Cauchy, these contributions to the integral 
vanish. Now we consider the non-zero contributions. 

Firstly we unravel the familiar case: $0 < \kplus < \Delta^{+}$ which gives a contribution $2 \pi i \Res (k_{c}^{-})$. At this point, 
however, the region appears far from familiar. The ratio $\kplus/P^{+}$ is restricted to $(0,1)$.  To relate with our previous notation, 
we must define $\kplus/\Pplus \equiv  Y = \frac{x - \xi}{1 - \xi}$ which restricts $x \in (\xi,1)$. 

Using the definition established in equation \ref{decomp}, we merely strip
away the $\kplus$ integral as well as the coupling to uncover
\begin{multline} \label{new}
\ndpd = \int \frac{d\kperp}{2 (2 \pi)^3}  \frac{1}{Y (Y+\Omega) (1-Y)} \frac{\Gamma(Y, \kperp - Y \Delta^{\perp}/2)}
{ 2 \Pplus P^{-} - \frac{m^2 +  \kperp{}^2 }{Y} - \frac{m^2 + (\kperp - \Delta^{\perp}/2)^2}{1 - Y}}  \\
\times \frac{\Gamma^{*} \Big( \frac{Y+\Omega}{1+\Omega}, \kperp + \big( \frac{1}{2} \frac{Y + \Omega}{1 + \Omega} -1 \big) \Delta^{\perp} 
\Big)}{ 2\Pplus (P^{-} - \Delta^{-}) - \frac{m^2 + (\kperp - \Delta^{\perp})^2}{Y + \Omega} - \frac{m^2 + (\kperp - \Delta^{\perp}/2)^2}
{1 - Y}}.
\end{multline} 
Simple algebra gives the relation $\frac{Y + \Omega}{1+ \Omega} = \frac{x+\xi}{1+\xi} = X$. As before we can use $\Delta^{-} = 
\frac{t + \Delta^{\perp}{}^2}{2 \Delta^{+}}$ to write $2 \Pplus \Delta^{-} = -(
t + \Delta^{\perp}{}^2)/\Omega$. (Back in section \ref{TD}, this relation was strikingly similar $2 \Pplus \Delta^{-} =
 - (t + \Delta^{\perp}{}^2)/\zeta$.)  We then exploit $\Delta \cdot \bar{P} = 0$
to find
\begin{equation}
- t = \frac{(1 + \Omega/2)^2 \Delta^{\perp}{}^2 + \Omega^2 M^2}{1 + \Omega}, 
\end{equation}
which relates to the expression earlier by $\Omega \to - \zeta$. Using the above, we can write $2 \Pplus \Delta^{-} = 
\Omega (M^2 + \frac{1}{4}\Delta^{\perp}{}^2)/(1 + \Omega)$. (Back in section \ref{TD}, we used the relation $2 \Pplus \Delta^{-} = 
\zeta (M^2 + \frac{1}{4}\Delta^{\perp}{}^2)/(1 - \zeta)$.)

Identification of the initial meson's wave function in Eq. \ref{new} is 
simple
\begin{equation}
\frac{\Gamma(Y, \kperp - Y \Delta^{\perp}/2)}{2\Pplus P^{-} - \frac{m^2 + \kperp{}^2}{Y} - 
\frac{m^2 + (\kperp - \Delta^{\perp}/2)^2}{1 - Y} }  = \psi(Y, \kperp - Y \Delta^{\perp}/2).
\end{equation}
The final meson's wave function requires us to manipulate the energy denominator
\begin{equation}
2 \Pplus (P^{-} - \Delta^{-}) - \frac{m^2 + (\kperp - \Delta^{\perp})^2}{Y + \Omega} - \frac{m^2 + (\kperp - \Delta^{\perp}/2)^2}{1 - Y} 
= \frac{1}{1 + \Omega} \Big( M^2 - \frac{m^2 + (\kperp + (X/2 - 1)\Delta^{\perp})^2}{X(1-X)} \Big).
\end{equation}
Upon shifting the $\kperp$ integral, we arrive at
\begin{equation}
\ndpd = \int \frac{d\kperp}{2 (2 \pi)^3} \frac{ 1 + \Omega}{Y (Y+ \Omega)(1-Y)} \psi(Y, \kperp + ( 1- \bar{x})
\Delta^{\perp} ) \psi^{*}(X, \kperp). 
\end{equation}
The pre-factors along with the conversion factor $1-\xi$ 
\begin{equation}
(1 - \xi) \frac{1 + \Omega}{Y(Y+ \Omega)(1-Y)} = \frac{1 - \xi^2}{(x^2 - \xi^2) (1-x)}
\end{equation}
identically line up to our expression \ref{ofpd1} and thus we see why the na\"{\i}ve replacement $\xi \to -\xi$ was justified. 

Lastly we consider the region $\Delta^{+} < \kplus < 0$. Given our identification of $Y$ as $\kplus / \Pplus$, we see $x$ is 
restricted to $(-\xi,\xi)$. To evaluate the integral Eq. \ref{tri2} in this r\'egime, 
we calculate $- 2 \pi i \Res (k_{a}^{-})$. From this expression, we toss away the $\kplus$ integral to uncover
\begin{multline} \label{new2}
\ndpd = \int \frac{d\kperp}{2 (2 \pi)^3} \frac{1}{Y(Y+\Omega)(1 - Y)} \frac{\Gamma
(Y, \kperp - Y \Delta^{\perp}/2)}{ 2 \Pplus \Delta^{-}  - 
\frac{m^2 + \kperp{}^2}{Y} + \frac{m^2 + (\kperp - \Delta^{\perp})^2}{Y + \Omega} } \\
\times \frac{\Gamma^{*}(X, \kperp + (X/2 - 1)\Delta^{\perp} )}{   2 \Pplus (P^{-} - \Delta^{-}) - 
\frac{m^2 + (\kperp - \Delta^{\perp})^2}{Y + \Omega} - \frac{m^2 + (\kperp - \Delta^{\perp}/2)^2}{1 -Y}    }.
\end{multline}
Comparing with our above equation (\ref{new}) the final state meson vertex and energy denominator are unchanged and can thus be
replaced by $(1 + \Omega) \psi^{*}( X, \kperp + (X/2 - 1) \Delta^{\perp} )$.

A little algebra needs to be employed to see
\begin{align}
2 \Pplus \Delta^{-}  - \frac{m^2 + \kperp{}^2}{Y} + \frac{m^2 + (\kperp - \Delta^{\perp})^2}{Y + \Omega} &=  - \frac{1}{\Omega} \Big( t + 
\frac{m^2 + (\kperp + \frac{Y}{\Omega}\Delta^{\perp})^2}{\frac{Y}{\Omega}
(\frac{Y}{\Omega} - 1) } \Big) \notag \\
& = - \frac{1}{\Omega} \Big( t - \frac{m^2 + (\kperp + (\sigma - 1)\Delta^{\perp})^2}{\sigma(1-\sigma)}   \Big),
\end{align}
where we used the relation $Y/\Omega = X/\zeta - 1 =  \sigma - 1$. 

Now in Eq. \ref{new2}, the initial meson's vertex can be related to a 
wave function only by using the interaction's crossing symmetry (since $Y<0$). Writing the remaining vertex in terms of the interaction,
as well shifting the $\kperp$ integral by $\Delta^{\perp}$ we see
\begin{multline}
\ndpd = \frac{- \Omega ( 1+ \Omega)}{Y(Y + \Omega)(1 - Y)} \int \frac{d\kperp dZ d\kaperp}{[2 (2 \pi)^3]^2 Z(1-Z)} 
\psi(Z, \kaperp) \\
\times V \Big( Z, \kaperp; Y, \kperp + (1 - Y/2) \Delta^{\perp} \Big)
\dw (\sigma, \kperp + \sigma \Delta^{\perp} | t ) \psi^{*}(X, \kperp + X \Delta^{\perp}/2).
\end{multline}   
For suspense, we have saved the pre-factors (and conversion factor) till the very end
\begin{equation}
- (1 - \xi) \frac{\Omega ( 1+ \Omega)}{Y(Y + \Omega)(1 - Y)} = - \frac{2 \xi (1 + \xi)}{(x^2 - \xi^2) (1-x)}.
\end{equation}
The $\xi$ symmetry of equation \ref{ofpd2} has been demonstrated and thus the covariant approach naturally incorporates the
subtlety of switching the $q\bar{q}$ pair annihilation from the initial to final state when $\xi \to -\xi$. This algebra
can easily be extended once interactions have been added at the photon vertex. 

\section*{Appendix B: SPD's, form factors and quark distributions}
Above we have made reference to verifying the sum rule Eq. \eqref{sum} without specifying what structure the form factor has. 
Here we connect our expressions for SPD's to the celebrated Drell-Yan-West formula for the form factor \cite{Drell:1970km} and
then analyze what neglecting the energy dependence of the $qq$ interaction has done to SPD's, form factors and quark distributions.  

Since Lorentz covariance constrains the $X$ integral of the SPD's to be independent of $\zeta$, we may find the electromagnetic 
form factor by integrating the SPD's in the limit of zero skewness ($\zeta \to 0$). This corresponds to using the highly simplifying 
Drell-Yan-West frame in which the plus component of the photon's momentum vanishes (here $\Delta^{+}= 0$). This limit is now 
complicated by possible singularities arising from the non-valence region ($X<\zeta$), like those encountered in \cite{pairs}. 
Ignoring this complication for the moment, we see the contribution from the valence region ($X>\zeta$) gives
\begin{equation} \label{dyw}
\lim_{\zeta \to 0} \int_{\zeta}^{1} X \mathcal{F}_{a}(X, \zeta, t) dX = \int \frac{dX d\kperp \theta[X(1-X)]}{2(2\pi)^3 X(1-X)} 
\psi(X, \kperp) \psi^{*} \big( X, \kperp + (1-X) \Delta^{\perp} \big)
\end{equation}
the Drell-Yan-West formula. 

It is interesting to first consider the non-valence contribution to the form factor coming from the bare-coupling SPD (Eq. \ref{ndpd2}). 
As this region extends from $X = 0$ to $\zeta$, to avoid singular momentum fractions, we change variables from $X$ to $\sigma = X/\zeta$ 
and then take the limit in which $\zeta$ vanishes. The contribution to the form factor reads
\begin{multline}
\lim_{\zeta \to 0} \int_{0}^{\zeta} X \ndpd dX  \sim  \zeta \; \int \frac{d\sigma \; d\kperp dZ d\kaperp}{[2 (2\pi)^3]^2 (1-\sigma) Z(1-Z)} 
\psi(\zeta \; \sigma,\kperp)\\
\times \bigdw \Big(\sigma, \kperp + \sigma \Delta^{\perp} \Big| t \Big) V \Big(Z, \kaperp; \zeta(\sigma - 1), \kperp + \Delta^{\perp} \Big) 
\psi^{*}(Z,\kaperp)
\end{multline}
which vanishes as $\zeta \; \psi(\zeta \; \sigma,\ldots)$ as $\zeta \to 0$. 
Thus for the Born approximation to the photon vertex, the Drell-Yan-West
formula indeed gives the form factor. 

We can, of course, employ the same change of variables technique to consider the contributions to the electromagnetic form factor 
from the true SPD's of section \ref{AGAIN}. The contribution from $\mathcal{F}_{b}(X,\zeta,t)$ Eq. \eqref{Fb} proceeds very similarly 
to the analysis for the bare-coupling SPD above. Additionally we must change variables from $W \in (0,\zeta)$ to $\omega = W/\zeta$ 
but in the end $\int X \; \mathcal{F}_{b} \; dX \sim \zeta \; \psi(\zeta \; \omega, \ldots)$ as $\zeta \to 0$. 

Lastly we consider the possible contribution from $\mathcal{F}_{c}(X, \zeta,t)$ Eq. \eqref{Fc}. The change of variables from $X$ 
to $\zeta$ runs as above. On the other hand, $W \in (\zeta, 1)$ giving us no reason for an addition change to $\omega$.  
Taking $\zeta$ to zero results in $\int X \; \mathcal{F}_{c} \; dX \sim \zeta \; 
V \Big(\frac{W}{\zeta},\frac{W}{\zeta}\Delta^{\perp};\ldots\Big)$ 
where $V$ is the crossed potential defined for a momentum fraction greater than one. Using the crossed potential for the 
Wick-Cutkosky model, this contribution to the form factor vanishes as $\zeta^2$. 

We have shown that both contributions $b$ and $c$ vanish in the limit of zero skewness. Within the impulse approximation, 
the Drell-Yan-West formula for the form factor is exact for our model. Thus we have seen, whether or not we include interactions 
at the photon vertex our model has the same form factor. The procedure here can be straightforwardly applied to the SPD's 
in the approximation schemes of section \ref{APPROX}. For precisely the same reasons above, both schemes preserve 
the Drell-Yan-West formula in the forward limit. 

Having demonstrated there are no singularities associated with the limit $\zeta \to 0$, we can also extract our model's quark distribution
function $q(x) \equiv F(x,0,0)$. This is then simply the forward limit of Eq. \ref{ndpd1}, namely
\begin{equation}
q(x) = \int \frac{d\kperp}{2(2\pi)^3 x(1-x)} |\psi(x,\kperp)|^2
\end{equation}
the square of the meson's wave function. In light of our above analysis, we know that the wave function vanishes quadratically at
the end points and thus $q(x)$ vanishes there as well. This is physical at $x = 1$ since one parton cannot posses all of the meson's 
plus momentum. At $x=0$, however, the non-valence partons can rescue one parton from carrying all the plus momentum and so such 
configurations have non-zero probability, i.e. $q(0) \neq 0$. Indeed the fact that our model distributions vanish at $x = 0$ is tied
directly to neglecting the mass squared dependence of the effective interaction in the two-body subspace. The non-valence 
probability amplitude $\langle \psi_{\text{nv}}| \psi_{\text{nv}} \rangle$ is simply
\begin{equation}
\langle \psi_{\text{nv}}| \psi_{\text{nv}} \rangle = - \frac{\partial}{\partial M^2} \langle \psi_{2} |V| \psi_{2} \rangle,
\end{equation}
where $\langle \psi_{2} |V| \psi_{2} \rangle$ is the projection of the interaction on the two-body subspace (simply denoted as $V$ above). 

Accounting for the mass squared dependence of the effective interaction allows for non-valence contributions to the quark distribution
functions. There are further consequences, however, given the unifying role of SPD's. The Drell-Yan-West formula, for example, will
have corrections (of particular importance to the form factor's $x$-integrand at small $x$), and adding interactions at the photon vertex
will then change the form factor. It has already been noticed \cite{Jason} that model deuteron wave functions have better covariance 
properties than their form factors calculated from Drell-Yan-West expressions. 
Lastly the most general consequence is to affect the SPD's, particularly at the crossover, e.g. 
$F(\zeta,\zeta,t) = 0$ seen here (as well as in \cite{Diehl:1999kh,Burkardt:2000uu}) stems directly from neglecting the mass squared
dependence. The crossover point thus probes a convolution of non-valence states for which this approach may possibly be apt in describing, 
however, it would require a reworking of the photon vertex that does not make use of the potential model's Green's function. 
(As for the light-cone Fock space expansion, calculation of wave functions at small $x$ is rather complicated but contains rich
information about the relation between different Fock components \cite{Antonuccio:1997tw}.)

\section*{Appendix C: GDA's and the time-like form factor}

Time-like form factors have been often avoided in the light-front approach due to the unavoidable presence of non-wave function 
vertices. This complication is mirrored in the scheme of \cite{Diehl:2001xz}: the time-like pion form factor does not have a direct 
decomposition in terms of pion Fock component overlaps alone. Here we show how the above analysis can be applied to time-like 
form factors; that is, we obtain expressions in terms of two-body Bethe-Salpeter wave functions and the four-point Green's functions.
Furthermore, we make the connection with the generalized distribution amplitude for our model which encodes 
the non-perturbative physics of two pion production \cite{Diehl:1998dk,Polyakov:1999td}. We attempt below to preserve the 
notation set up in these references. 

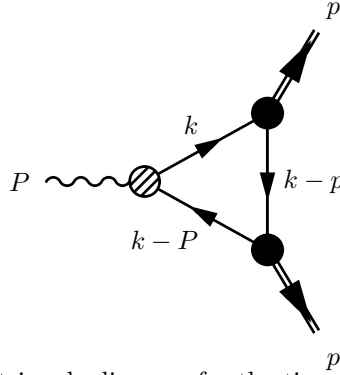
\begin{figure}
\begin{center}
\begin{fmffile}{fmftimetri}
\begin{fmfchar*}(40,40)
    \fmfleft{a,b,c}  \fmflabel{$P$}{b}
    \fmf{photon}{b,bee}
    \fmf{fermion,tension=.4,label=$k$,label.side=left}{bee,Zee}
    \fmf{fermion,tension=.4,label=$k-p$,label.side=left}{Zee,Zam}	
    \fmf{fermion,tension=.4,label=$k- P$,label.side=left}{Zam,bee}
		
    \fmf{heavy}{Zee,fc}
    \fmf{heavy}{Zam,fa}
    \fmfright{fa,fb,fc} 
    \fmflabel{$p$}{fc}
    \fmflabel{$p^{\prime}$}{fa}	
    \fmfv{decor.shape=circle,decor.filled=full,decor.size=.1w}{Zee,Zam}
    \fmfv{decor.shape=circle,decor.filled=shaded,decor.size=.1w}{bee}	
\end{fmfchar*}
\end{fmffile}
\caption{The triangle diagram for the time-like, pion form factor}
\label{timetri}
\end{center}
\end{figure}

The time-like pion form factor $F(s)$ is defined by (see Figure \ref{timetri})
\begin{equation}
\langle \pi(p) \; \pi(\pp) \;| \;J^{\mu}(0)\; | \;0 \;\rangle = (p - \pp)^{\mu} F(s),
\end{equation}  
where $s = (p + \pp)^2$ is the center of mass energy squared. Here the pion is merely our $qq$ pair. 
Now define $P^{\mu} = p^{\mu} + \pp^{\mu}$ and $\zeta = \pplus / \Pplus$. We can work out the
kinematics of this reaction to find
\begin{align}
P^{-}      & = \frac{s}{2 \Pplus}          \notag \\ 
p^{-}      & = \frac{(1-\zeta)s}{2 \Pplus} \notag \\
\pperp{}^2 & = s (1 - \zeta) \zeta - M^2,
\end{align}
where $M$ is the pion mass and we work in the frame in which $\mathbf{P}^{\perp} = 0$. . 

The GDA for our model has a definition in terms of a non-diagonal matrix element of bi-local field operators
\begin{equation}
\GDA =  \int \frac{dx^{-} e^{i z \Pplus x^{-}}} {2 \pi (2 z - 1)} \langle \; \pi(p) \; \pi(\pp) \; | \;  
q(x^{-}) \overset{\leftrightarrow}\partial {}^{+} q(0) \;  | 0  \; \rangle.
\end{equation}
Such a definition of the GDA leads us immediately to the sum rule
\begin{equation}
\int \frac{2 z -1}{2 \zeta - 1} \GDA dz = F(s),
\end{equation}
and hence a means to calculate $\Phi$ from the integrand of the time-like form factor. 

Returning to Figure \ref{timetri}, we can write down the expression for the time-like form factor
\begin{equation}
- i \int \frac{d^4 k}{(2 \pi)^4} \Gamma^{+}(k,P) G_{o}(k) \Gamma(k,p) G_{o}(k-p) \Gamma(k-p,P-p) G_{o}(k-P)
\end{equation}
and perform the light-front reduction by integrating out the minus dependence. Whenever troublesome momentum fractions appear, 
we use the equation of motion \eqref{wein} to insert the crossed interaction. Lastly we merely extract the derivative coupling 
and $z$-integral from the photon vertex function to uncover the GDA. Using $w \equiv \kplus/\Pplus$, the relevant 
contribution from the photon vertex appears as
\begin{equation}
\delta(q^{+} - z \Pplus) \frac{i \Gamma^{+}(k,P)}{(2 z -1) \Pplus} \bigdw (w, \kperp | s) = 
\int \frac{d\qperp}{2 (2 \pi)^3 z (1-z)} G(z,\qperp; w, \kperp | s),
\end{equation} 
with $z \in (0,1)$. 

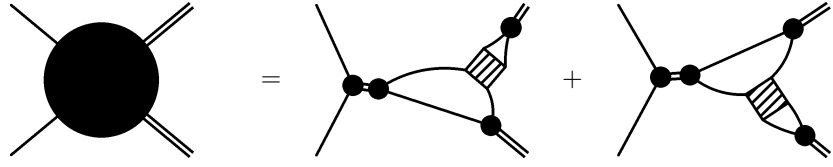
\begin{figure}
\begin{center}
\begin{fmffile}{fGDA}
	\parbox{35mm}{
	\begin{fmfgraph}(30,20)
	\fmfleft{a,c}
	\fmf{plain,tension=.2}{a,in}
	\fmf{plain,tension=.2}{c,in}
	\fmf{double,tension=.2}{in,b}
	\fmf{double,tension=.2}{in,d}
	\fmfv{decor.shape=circle,decor.filled=full,decor.size=.5w}{in}
	\fmfright{b,d}
	\end{fmfgraph}} $=$ 	
	\parbox{35mm}{
	\begin{fmfgraph}(35,20)
	\fmfleft{a,c}  
    	\fmf{plain,tension=.7}{a,in}
	\fmf{plain,tension=.7}{c,in}
    	\fmf{double,tension=2}{in,out}
	\fmf{plain,tension=.3}{out,Zee}
     	\fmfpolyn{shade,tension=.3}{Z}{4}
	\fmf{plain,tension=.2,left=.2}{out,Z3}
        \fmf{plain,tension=.5,right=.2}{Zee,Z4}
	\fmf{plain,tension=.3,left=.1}{Z1,Zam}
    	\fmf{plain,tension=.5,right=.1}{Z2,Zam}
 	\fmf{double}{Zam,fc}
    	\fmf{double}{Zee,fa}
    	\fmfright{fa,space,fc} 
    	\fmfv{decor.shape=circle,decor.filled=full,decor.size=.07w}{in,out,Zee,Zam}
    	\end{fmfgraph}} $+$
	\parbox{35mm}{
	\begin{fmfgraph}(35,20)
	\fmfleft{a,c}  
    	\fmf{plain,tension=.7}{a,in}
	\fmf{plain,tension=.7}{c,in}
    	\fmf{double,tension=2}{in,out}
	\fmf{plain,tension=.3}{out,Zee}
     	\fmfpolyn{shade,tension=.3}{Z}{4}
	\fmf{plain,tension=.5,right=.2}{out,Z4}
        \fmf{plain,tension=.2,left=.2}{Zee,Z3}
	\fmf{plain,tension=.4,left=.1}{Z2,Zam}
    	\fmf{plain,tension=.4,right=.1}{Z1,Zam}
 	\fmf{double}{Zee,fc}
    	\fmf{double}{Zam,fa}
    	\fmfright{fa,space,fc} 
    	\fmfv{decor.shape=circle,decor.filled=full,decor.size=.07w}{in,out,Zee,Zam}
    	\end{fmfgraph}}  	

\end{fmffile}
\end{center}
\caption{Graphical representation of contributions to the GDA Eq. (\ref{gda}): $\Phi = \phi_{a} + \phi_{b}$. The shaded box again represents
the crossed interaction and the internal double line represents the four-point Green's function.}
\label{GDAfig}
\end{figure}

Carrying out this evaluation yields
\begin{multline}   \label{gda}
\GDA = \int \frac{d\qperp dw d\kperp dx d\kaperp}{ [2 (2 \pi)^3 ]^3 w (w-\zeta) (1-w) x (1-x) z (1-z)} \psi(x,\kaperp) 
G(z,\qperp ; w ,\kperp | s )\\ 
\times \Bigg( \phi_{a}(w, \kperp; x, \kaperp) \; \theta [ w (\zeta - w) ] + \phi_{b}(w,\kperp;x,\kaperp) \; \theta[(1-w)(w - \zeta)] \Bigg) 
\end{multline}
where we have abbreviated
\begin{align} \label{phis}
\phi_{a}(w,\kperp;x,\kaperp) & = \zeta \; V \Big( x, \kaperp; y, \kperp + (1-y)\pperp \Big) 
\psi \Big(\frac{w}{\zeta}, \kperp - \frac{w}{\zeta} \pperp  \Big) \notag \\
\phi_{b}(w,\kperp;x,\kaperp) & = (\zeta - 1) \; V \Big( x, \kaperp; \frac{w}{\zeta}, \kperp - \frac{w}{\zeta} \pperp \Big) 
\psi \Big( y, \kperp + (1-y) \pperp \Big),
\end{align}
with $y = \frac{w - \zeta}{1-\zeta}$. The two contributions ($a$ and $b$) to the GDA can be interpreted graphically, see Figure
\ref{GDAfig}.

We can furthermore decompose the GDA in the approximation schemes set forth in section \ref{APPROX}. In the 
non-relativistic scheme, we use the approximation to the Green's function in Eq. \ref{nrgreen} to derive
\begin{equation}
\GDA^{\text{NR}} = \Bigg( \phi_{a}^{\text{NR}}(\zeta,s) \; \theta \big[\zeta - \frac{1}{2} \big] + \phi_{b}^{\text{NR}}(\zeta,s) \; 
\theta \big[\frac{1}{2} - \zeta \big]   \Bigg) 
\int \frac{d\qperp}{2(2\pi)^3 z (1-z)} I(z, \qperp | s ) \dw (z, \qperp | s ),
\end{equation} 
where
\begin{align}
\phi_{a}^{\text{NR}}(\zeta,s) & = \frac{\zeta}{\frac{1}{2} - \zeta} \psi(\mathbf{r}=0) 
V \Big(\frac{1}{2}, \mathbf{0}^{\perp}; \frac{\frac{1}{2} - \zeta}{1 - \zeta}, - \frac{\pperp/2}{1 - \zeta} \Big) 
\psi \Big(\frac{1}{2 \zeta}, -\frac{\pperp}{2\zeta} \Big) \notag \\
\phi_{b}^{\text{NR}}(\zeta,s) & = - \frac{1-\zeta}{\frac{1}{2} - \zeta} \psi(\mathbf{r}=0)  
V \Big(\frac{1}{2}, \mathbf{0}^{\perp}; \frac{1}{2 \zeta}, -\frac{\pperp}{2 \zeta}\Big) 
\psi \Big( \frac{\frac{1}{2} - \zeta}{1-\zeta}, -\frac{\pperp/2}{1 - \zeta} \Big),
\end{align}
where $\psi(\mathbf{r}=0)$ is defined in Eq. \ref{origin}.
In this form, we see the non-relativistic approximation maintains continuity at $\zeta = 1/2$. Apart from $\zeta$ dependent
coefficients, the structure of the non-relativistic GDA is identical to that of the non-relativistic SPD in the non-valence region 
(with $\sigma$ replaced with $z$), see Eq. \eqref{NRNV}. 

Lastly appealing to the closure approximation for the Green's function, we arrive at the approximate GDA
\begin{multline}
\GDA^{\text{CL}} = \int \frac{d\qperp dx d\kaperp}{[2 (2 \pi)^3]^3 z (1-z)(z - \zeta)x(1-x)} \psi(x, \kaperp) \\ 
\times \Bigg( \frac{\phi_{a}(z,\qperp; x, \kaperp)} {s - \mn_{a}} \theta[z(\zeta - z)]  +  \frac{\phi_{b}(z,\qperp; x, \kaperp)}
{s - \mn_{b}} \theta[(1 - z)(z - \zeta)] \Bigg). 
\end{multline}
Notice given the form of the functions \ref{phis}, the closure approximation to the GDA is continuous at $z = \zeta$. 
This is a curious fact (in contrast to SPD's) since the factorization theorem for GDA's does not depend on continuity at 
$z = \zeta$ \cite{Polyakov:1999td}. Additionally at large $s$, we can apply the Born approximation to the Green's function. 
The resulting GDA's are identical to bare-coupling GDA's, which is a further contrast to SPD's.


\begin{thebibliography}{99}
\bibitem{boost}
Choosing now the symmetrical frame, the transverse momentum $\Delta^{\perp}$ in section \ref{TD} is related 
to that in section \ref{DVCS} by a boost: $\Delta^{\perp} \Big|_{\text{sym}} = ( 1 + \xi) \Delta^{\perp} \Big|_{\text{asym}}$.
\bibitem{comment}
In our previous work, we did not remove the derivative coupling at the photon vertex from the SPD. Thus we differ by a factor
of $x$ in the numerator.
\bibitem{alt}
Alternatively, one could appeal to Eq. \ref{LS} to deal with the problematic momentum fractions. In this scheme, we insert 
the T-matrix instead of the Green's function in the photon vertex function. The problematic momentum fraction gives rise 
to a crossed potential as well as a crossed T-matrix. To unravel the SPD's in terms of two-body wave functions one utilizes 
the two potential theorem, and the final results are, of course, the same.
\end{thebibliography}
\end{document}